\documentclass{article}

\PassOptionsToPackage{numbers,sort&compress}{natbib}
\usepackage[preprint]{neurips_2026}
\usepackage[utf8]{inputenc}
\usepackage[T1]{fontenc}
\usepackage[hypertexnames=false]{hyperref}
\usepackage{url}
\usepackage{booktabs}
\usepackage{amsmath,amsfonts}
\usepackage{graphicx}
\usepackage{array}
\usepackage{algorithm}
\usepackage{algorithmic}
\usepackage{listings}
\usepackage{microtype}
\usepackage{xcolor}

\hypersetup{
  pdftitle={Inferring Value Criteria from Ordinal Preferences: An Iterative In-Context Learning Framework for Music Generation},
  pdfauthor={Futa Hidaka, Naomi Imasato, Kazuki Miyazawa, and Takato Horii},
  pdfkeywords={in-context learning, preference learning, symbolic music generation, large language models, personalization, computational creativity}
}

\newcommand{\vflabel}[1]{\textit{#1}}
\newcommand{\vfcomb}[2]{\vflabel{#1}~\ensuremath{\times}~\vflabel{#2}}

\title{Inferring Value Criteria from Ordinal Preferences: \\An Iterative In-Context Learning Framework for Music Generation}
\author{%
  Futa Hidaka$^{1}$ \quad Naomi Imasato$^{1}$ \quad
  Kazuki Miyazawa$^{1}$ \quad Takato Horii$^{1,2}$ \\
  $^{1}$The University of Osaka \quad $^{2}$The University of Tokyo \\
  \texttt{u513279d@ecs.osaka-u.ac.jp}
}

\begin{document}

\maketitle

\begin{abstract}
Adapting a generative music system to an individual's taste requires learning what that listener values. Listeners can rank pieces, but their underlying criteria may be tacit and difficult to articulate. We ask whether and under what conditions a large language model (LLM) can adapt symbolic music generation from rankings alone and construct transferable natural-language descriptions of value criteria. In our iterative in-context learning framework, the LLM formulates hypotheses, generates candidate pieces in ABC notation, receives a ranking, and periodically infers and verbalizes value criteria from history to guide later generation. We evaluate the framework against 16 simulated raters in 480 adaptation runs using mixed-effects modeling, an ablation, and transfer tests on unseen music. Overall, the framework did not outperform a feedback-free diverse-generation baseline, but did so for two value functions with targets difficult to reach through simple sampling. How atypical the target was relative to the LLM's feedback-free generation tendencies predicted adaptation difficulty. Moreover, higher value during adaptation did not imply identification of the criterion as a general rule. On unseen music, acquired descriptions and histories improved generation for more value functions than they improved preference prediction, which remained near chance. Some gains were associated with acoustic proximity to music in the context, but others were not. These findings show that rankings alone can guide generation under limited conditions, while transferable criterion inference remains constrained by the foundation model's ability to recognize, reason about, and verbalize musical attributes.
\end{abstract}


\section{Introduction}\label{sec:intro}
Computational creativity commonly characterizes creative people and artifacts in terms of novelty and value \citep{sternberg1999handbook}, and empirical criteria have been proposed to operationalize them \citep{ritchie2007some}. Just as novelty depends on what an artifact is compared against \citep{boden2004creative}, value depends on who evaluates it and in what context. A generated artifact must therefore differ from existing data and be recognized as valuable. \citet{tokui2020can} incorporated novelty into generation by using a Generative Adversarial Network (GAN) to produce rhythm patterns not classified as existing genres. \citet{fujimoto2019generative} proposed a GAN-based framework that learns from evaluator-valued data while expanding the generative space through entropy maximization. However, these studies provide only limited mechanisms for turning evaluations into a sequential search policy, leaving raters to evaluate many candidates.

Treating value as evaluator-dependent raises the question of whose value to learn. Music-generation methods have learned reward models from large-scale pairwise user preferences and fine-tuned generators through Reinforcement Learning from Human Feedback \citep{cideron2024musicrl}; others perform Direct Preference Optimization (DPO) on preference pairs constructed with automated aesthetic models \citep{chen2025diffrhythm}. The resulting models need not match any particular individual's preferences. Indeed, musical pleasure depends on the balance between predictability and uncertainty, and the optimal balance differs substantially across individuals. This individual difference showed moderate temporal stability and was not explained by musical training or self-reported exposure to jazz \citep{mas2025predictive}. In judgments of music similarity, when salient cues such as genre were unavailable and judgments had to rely on finer musical features, both inter- and intra-rater agreement decreased, although intra-rater agreement remained higher \citep{flexer2021evaluation}. These findings motivate individual-level adaptation rather than relying solely on aggregated preferences, while cautioning that individual judgments are not perfectly stable.

Human-in-the-loop optimization has pursued individual adaptation by iteratively improving generation from evaluations. Recent approaches use the in-context learning (ICL) capabilities of large language models (LLMs) to update generation policies from interaction histories. Some require detailed textual feedback explaining the rater's choices and suggesting improvements \citep{lee2025feedback}, whereas others attempt adaptation from preference signals alone \citep{yu2024icpl}. The former assumes that raters can articulate their reasons. The latter avoids this requirement, but preference signals provide limited information, and noise or ambiguity may degrade adaptation. The former's assumption that raters can verbalize their reasons has psychological limitations. Some criteria underlying preferences may be tacit \citep{polanyi1966tacit}; people may lack direct introspective access to their judgment processes and construct reason reports from plausible causal theories \citep{nisbett1977telling}. Analyzing reasons can itself alter which criteria receive attention \citep{wilson1991thinking}. Preferences may also be constructed and updated through choice rather than fully determined beforehand \citep{ariely2008actions}. Thus, verbal reports and preference signals impose different constraints, and adaptation from limited evaluator information remains challenging. Despite their limited information and potential noise, preference signals may provide feedback for tracking preference patterns while avoiding distortions introduced by verbalization.

\begin{figure}[t]
  \centering
  \includegraphics[width=\columnwidth]{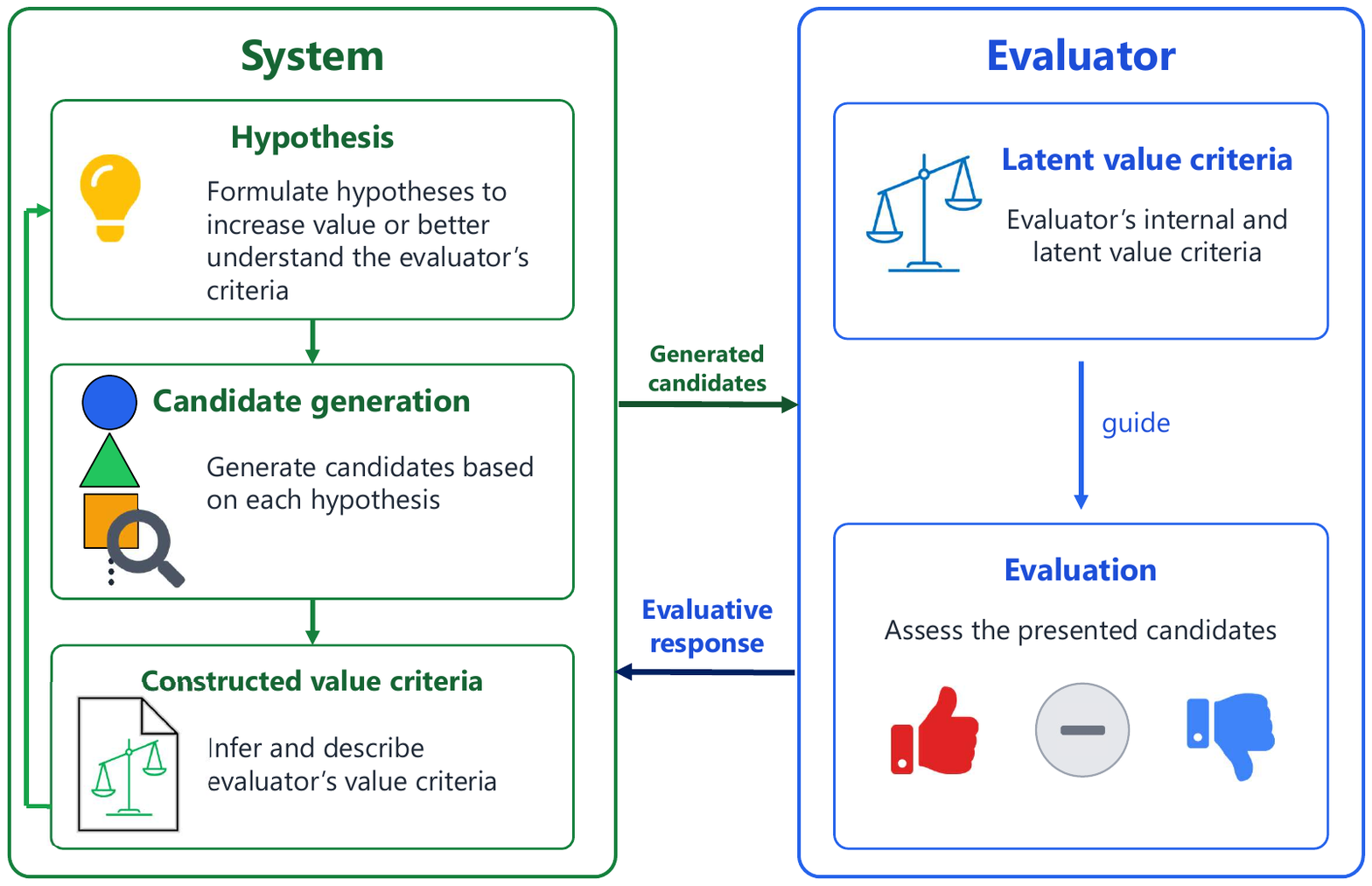}
  \caption{Value learning from observed evaluations: a working hypothesis of the criteria is constructed and reused in generation.}
  \label{fig:value_learning_concept}
\end{figure}

We focus on the value component of creativity in music generation. Our setting comprises a system that generates music and a rater who evaluates it. We define \emph{value} as the desirability the rater assigns to a piece and \emph{value criteria} as the rater's latent rules for determining which musical attributes are desirable. Because the system cannot observe these criteria directly, generating music the rater values requires learning them through trial and error from observable feedback (Figure~\ref{fig:value_learning_concept}). We instantiate this conceptual framework in music generation: an LLM constructs the criteria as a working hypothesis from preferences without explanations and reuses it for generation. In each iteration, the LLM forms hypotheses and generates music, then receives the rater's ranking of multiple pieces (the \emph{preference signal}). The system periodically infers the criteria from the accumulated history and feeds the resulting verbalized working hypothesis (the \emph{value-criteria description}) into subsequent iterations. As a preliminary validation before studies with human raters, our experiments use simulated raters to evaluate adaptation from preference signals and transfer of the descriptions to unseen music.

We address three research questions. RQ1: Can the system iteratively adapt to a rater's hidden value criteria using only preference signals, without natural-language explanations? RQ2: Can the value-criteria descriptions and generation and ranking histories acquired during adaptation transfer beyond the music observed during adaptation, to preference prediction on unseen music and generation of new music? RQ3: What factors support or hinder adaptation and transfer? We contribute an LLM-based framework that constructs and reuses value-criteria descriptions from preference signals, together with controlled evaluations and analyses of adaptation and transfer using simulated raters.

\section{Related Work}\label{sec:related}
In music generation, evaluator feedback has been used in both online individual adaptation and offline preference alignment. The former includes GenJam's human-guided evolution of jazz solos \citep{biles1994genjam}, interactive differential evolution for sound synthesis \citep{wang2024interactive}, and Bayesian optimization of musical latent spaces \citep{zhou2020melody}. The latter includes MusicRL, trained from large-scale pairwise preferences \citep{cideron2024musicrl}, and DPO-based fine-tuning with an aesthetic model as a proxy evaluator \citep{chen2025diffrhythm}. Online methods can adapt to individuals but require continued evaluation. Offline alignment can use many preferences or automated scores, but the resulting model need not match any particular rater's preferences.

Inferring a rater's internal criteria from limited preference signals has been studied in preference-based reinforcement learning \citep{wirth2017survey} and active preference learning \citep{sadigh2017active}. These approaches commonly posit a latent utility function and infer it as weights over predefined features. OPEN uses an LLM to identify domain-specific features, but still represents the criteria as linear weights over them \citep{handa2024bayesian}. Feedback unexplained by an existing feature space may instead require learning the features themselves \citep{bobu2021feature}. This problem is pronounced in music, where perceptual attributes such as tonal stability and melodiousness can be difficult to define and capture with hand-crafted descriptors \citep{aljanaki2018data}. Because an LLM can serve as a proxy evaluator when given natural-language descriptions or examples \citep{kwon2023reward}, criteria may instead be retained as context. We use this flexibility to construct a natural-language working hypothesis directly from preferences.

LLMs are increasingly used as inference-time optimizers using histories and feedback \citep{yang2024opro}, and are beginning to support adaptation to human preferences. Feedback Descent uses binary preferences, explanations of why the winner is better, and improvement directions \citep{lee2025feedback}. In a molecular-optimization ablation, random or binary-only feedback degraded performance; in vector-graphics generation, five iterations of accumulated context generally outperformed direct prompting with stated criteria, suggesting the benefit of iterative context acquisition. In-Context Preference Learning (ICPL) instead iteratively improves reward functions from output preferences without rationales \citep{yu2024icpl}. Our framework adopts its loop but generates music and has the LLM infer and internally construct the rater's value criteria as a natural-language working hypothesis for generation.

Natural-language criteria have been extracted post hoc from pairwise-comparison datasets by Inverse Constitutional AI \citep{findeis2025inverse}, or refined interactively through free-form conversational preference elicitation \citep{li2025eliciting} and dialogue eliciting explanations for preferences \citep{blair2025reflective}. The latter two approaches rely on verbal reports. Methods avoiding such reports have inferred natural-language rules through active experiments \citep{piriyakulkij2024doing}, but identify rules with a single objectively verifiable answer from experimental outcomes. Although our controlled evaluation likewise uses predefined scoring rules with objectively verifiable outcomes, we treat each rule as a simulated rater's latent criterion and test whether a working hypothesis formed from rankings alone transfers to prediction and generation on unseen music.

\section{Proposed Method}\label{sec:method}
The proposed system follows ICPL's iterative structure \citep{yu2024icpl}: an LLM generates multiple candidates, and the rater's ranking is added to the context for subsequent generation. Unlike ICPL, it generates music rather than reward functions and adds hypothesis generation and value-criteria inference. The rater provides only a ranking of the new candidates and the current best piece, without explanations.

The framework can interface with text-to-music models, digital audio workstations, or programmatic tools; our implementation uses ABC notation, previously used for LLM-based music understanding and generation \citep{yuan2024chatmusician,zhou2024llmsreason}, and the open-weight general-purpose gpt-oss-120b \citep{agarwal2025gptoss}; its pinned revision aids reproducibility (Section~9).

\begin{figure}[!t]
\centering
\includegraphics[width=\columnwidth, trim=30pt 30pt 30pt 100pt, clip]{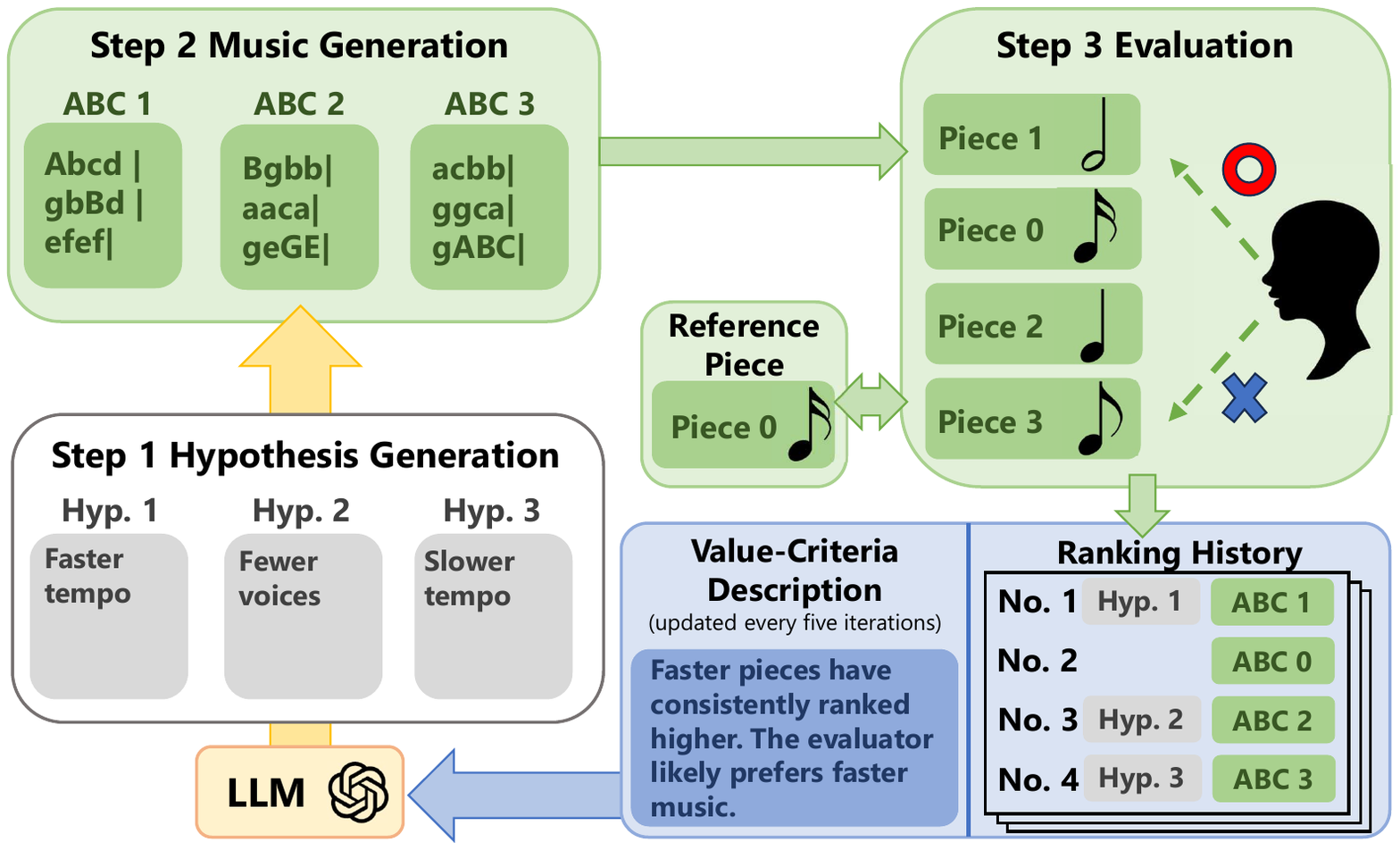}
\caption{Data flow of the proposed system.}
\label{fig:system_overview}
\end{figure}

The initial piece serves as the first reference piece and is thereafter replaced by the top-ranked piece from the preceding iteration. At each iteration, the LLM forms three hypotheses, generates one piece from each, and obtains a ranking of the three new pieces and the reference. Every five iterations, it constructs a value-criteria description from the ranking history. Figure~\ref{fig:system_overview} shows the data flow, Algorithm~\ref{alg:proposed_system} specifies the procedure, and Appendix A of the Supplementary Material provides the full prompts.

Each LLM call receives step-specific inputs plus the latest value-criteria description and ranking history. The history retains each iteration's hypotheses, ABC scores, and rankings.

\subsection{Hypothesis generation}
In the hypothesis-generation step (Figure~\ref{fig:system_overview}, bottom left; line~\ref{line:hypothesis} of Algorithm~\ref{alg:proposed_system}), the LLM generates three hypotheses about musical modifications that may either lead to a higher-rated piece or improve its understanding of the value criteria. Falsification-style hypotheses that may temporarily lower value are permitted when they support the latter aim. Retaining each hypothesis with its ABC score and ranking allows subsequent inference to relate the intent of a generation to its outcome.

\subsection{Music generation}
In the music-generation step (Figure~\ref{fig:system_overview}, top left; line~\ref{line:musicgeneration} of Algorithm~\ref{alg:proposed_system}), the LLM generates one ABC score from each hypothesis, yielding three pieces per iteration.

\subsection{Evaluation}
In the evaluation step (Figure~\ref{fig:system_overview}, top right; line~\ref{line:evaluation} of Algorithm~\ref{alg:proposed_system}), the rater ranks the three new pieces and the reference piece. Ties are marked as \texttt{(tied)}, with a new piece placed above the reference so that it replaces the reference. Ties among new pieces are resolved in generation order.

\subsection{Value-criteria inference and ranking-history management}
In the value-criteria inference step (Figure~\ref{fig:system_overview}, bottom right; line~\ref{line:valuecriteria} of Algorithm~\ref{alg:proposed_system}), every five iterations the LLM infers and verbalizes the rater's criteria from the ranking history and past descriptions. The prompt explicitly instructs the LLM to attend to the ABC scores themselves rather than confining its judgment to the natural-language materials such as the hypotheses and past descriptions. The resulting description immediately enters the shared context. When the history reaches 10 iterations, the oldest five are removed after inference; their information may persist indirectly through the latest description.

\begin{algorithm}[!t]
\caption{Proposed-system procedure}
\label{alg:proposed_system}
\begin{algorithmic}[1]
\renewcommand{\algorithmiccomment}[1]{~~// #1}
\REQUIRE Initial piece, number of iterations $N$, rater
\ENSURE Sequence of generated pieces and value-criteria description
\STATE reference piece $\leftarrow$ initial piece
\STATE ranking history $\leftarrow$ empty list,~ value-criteria description $\leftarrow$ empty
\FOR{$t = 1$ \TO $N$}
    \STATE shared context $\leftarrow$ (value-criteria description, ranking history)
    \STATE $h_1, h_2, h_3 \leftarrow \mathrm{LLM}($reference piece, shared context$)$ \COMMENT{hypothesis generation} \label{line:hypothesis}
    \FOR{$k = 1$ \TO $3$}
        \STATE $m_k \leftarrow \mathrm{LLM}(h_k$, reference piece, shared context$)$ \COMMENT{music generation} \label{line:musicgeneration}
    \ENDFOR
    \STATE ranking $\leftarrow$ rater$(m_1, m_2, m_3$, reference piece$)$ \COMMENT{evaluation} \label{line:evaluation}
    \STATE append (hypotheses, ABC scores, ranking) to ranking history
    \IF{$t \bmod 5 = 0$}
        \STATE value-criteria description $\leftarrow \mathrm{LLM}($ranking history, past value-criteria descriptions$)$ \COMMENT{value-criteria inference} \label{line:valuecriteria}
        \IF{ranking history has reached 10 iterations}
            \STATE delete the oldest five iterations from the ranking history
        \ENDIF
    \ENDIF
    \STATE reference piece $\leftarrow$ top-ranked piece
\ENDFOR
\end{algorithmic}
\end{algorithm}
\section{Experimental Setup}\label{sec:setup}
Experiment~1 evaluates adaptation through iterative generation, and Experiment~2 evaluates transfer of the acquired criteria to unseen music. To evaluate system performance quantitatively, we use 16 value functions implementing the raters and a shared dataset of 60 pieces.

\subsection{Value functions}
A value function scores music from 0 to 100, assigning 100 to its target state and lower values with increasing deviation. We selected 13 single and three pairwise-combination functions spanning distinct features and both symbolic and acoustic measures (Table~\ref{tab:value_functions}). During evaluation, the presented pieces are ranked by these values.

Of the single functions, the eight targeting pitch range, pitch, tempo, or duration map distance from the target through exponential decay, and the four contour and pitch-class functions use the proportion of elements satisfying the target. For the tempo functions, tempo is expressed in beats per minute (BPM). \vflabel{M2E-Sad} applies Music2Emo \citep{kang2025music2emo} to audio rendered with FluidSynth \citep{fluidsynth_2_2_5} and scales the \texttt{sad} probability to 0--100. Appendix B provides the full definitions.

Each combination function is defined as the geometric mean $\sqrt{v_a \cdot v_b}$ of its two constituents, strongly requiring simultaneous achievement of both targets. This tests adaptation to preferences involving multiple musical features.

\begin{table*}[!t]
\centering
\footnotesize
\begin{tabular}{lll}
\toprule
Category & Value function & Target state \\
\midrule
Pitch range & \vflabel{C5--B5 range} & All pitches within C5--B5 \\
            & \vflabel{C2--B2 range} & All pitches within C2--B2 \\
\addlinespace
Pitch & \vflabel{G4 pitch} & All pitches equal to G4 \\
      & \vflabel{B6 pitch} & All pitches equal to B6 \\
\addlinespace
Pitch contour & \vflabel{Ascending contour} & Ascending ratio of highest notes 100\% \\
              & \vflabel{Descending contour} & Descending ratio of highest notes 100\% \\
\addlinespace
Tempo & \vflabel{150~BPM tempo} & Tempo of 150~BPM \\
      & \vflabel{30~BPM tempo} & Tempo of 30~BPM \\
\addlinespace
Duration & \vflabel{60~s duration} & Duration of 60~s \\
         & \vflabel{5~s duration} & Duration of 5~s \\
\addlinespace
Pitch class & \vflabel{A locrian PCs} & In-set pitch-class ratio of 100\% \\
            & \vflabel{B major PCs} & In-set pitch-class ratio of 100\% \\
\addlinespace
Emotion & \vflabel{M2E-Sad} & Music2Emo \texttt{sad} probability of 1 \\
\addlinespace
Combination & \vfcomb{C5--B5 range}{M2E-Sad} & Simultaneous achievement of both constituents \\
            & \vfcomb{C2--B2 range}{Ascending contour} & Same as above \\
            & \vfcomb{Ascending contour}{150~BPM tempo} & Same as above \\
\bottomrule
\end{tabular}
\caption{Value functions and target states.}
\label{tab:value_functions}
\end{table*}

\subsection{Music dataset}
We mechanically selected 60 pieces from the OpenScore Lieder Corpus \citep{openscore_lieder_zenodo_v3,gotham_jonas_2022_openscore_lieder}. During MusicXML-to-ABC conversion, we removed lyrics, layout directives, and ornaments not represented in MIDI. To limit prompt length, we retained pieces that converted with abc2midi \citep{abcmidi_abc2midi_2022} and rendered with FluidSynth, had ABC scores of at most 3,000 bytes, lasted 10--100 seconds, and contained at least 50 MIDI note-on events. Thirty served as initial pieces for iterative generation and 30 as unseen test pieces in Experiment~2.

\section{Experiment 1: Iterative Adaptation from Ranking Preferences}\label{sec:exp1}

\subsection{Purpose and setup}
Experiment~1 tests whether the proposed system can adapt to simulated raters through iterative generation and evaluation. We compare adaptation with diverse generation, which uses no value-function evaluation history, and analyze adaptation factors and the contribution of value-criteria descriptions.

We evaluated all 480 cells comprising 16 value functions and 30 initial pieces. Each cell began with the initial piece as the reference at iteration 0; iterations 1--20 each generated three pieces from three hypotheses (60 total). As a comparison baseline, we included the diverse-generation condition to show how far value could reach when the same number of pieces were generated without any information derived from the value functions. For each initial piece, it generated one common set of 60 pieces sequentially (1,800 total); each prompt contained the initial piece, up to the 30 most recent ABC scores, and an instruction to generate a meaningfully different piece (Prompts S6 and S7). It omitted hypothesis generation, ranking feedback, and criteria descriptions, but used the same LLM, ABC-score conversion validation, and retry procedure. Each common set was evaluated with all 16 value functions. Both conditions involved 61 unique pieces, but differed structurally: twenty sequential four-piece rankings versus a global best-of-61 selection.

With $v_{f,m}^{(0)}$ the initial value of value function $f$ on initial piece $m$ and $v_{f,m}^{(t)}$ the value of the reference piece obtained by iteration $t$, the adaptation ratio $r_{f,m}^{(t)}$ is defined by Equation~\ref{eq:exp1_adaptation_ratio}:
\begin{equation}
r_{f,m}^{(t)}=
\frac{v_{f,m}^{(t)}-v_{f,m}^{(0)}}
{100-v_{f,m}^{(0)}}\times 100
\label{eq:exp1_adaptation_ratio}
\end{equation}
The adaptation ratio is the percentage of the initial-to-maximum gap closed by iteration $t$. Cells with an initial value of 100 would be excluded because the denominator is zero; none occurred, so all 480 were analyzed. Comparisons use iteration 20; for diverse generation, $v_{f,m}^{(t)}$ is the maximum over its 60 pieces and the initial piece.

Overall condition differences were tested with a linear mixed-effects model of adaptation ratio, with condition (reference: diverse generation) as a fixed effect and crossed random intercepts for initial piece and value function. Per function, paired one-sided $t$-tests matched on initial piece tested whether the proposed condition was higher; the 16 $p$-values were Holm-corrected. Per-function tests were one-sided because the directional hypothesis---that feedback-guided iteration outperforms feedback-free generation---was specified a priori. All mixed-effects models were estimated using restricted maximum likelihood; fixed-effect $p$-values and 95\% confidence intervals used the Wald normal approximation.

\subsection{Results}

\subsubsection{Trajectory of iterative adaptation and comparison with diverse generation}
\begin{figure*}[!t]
\centering
\includegraphics[width=\textwidth]{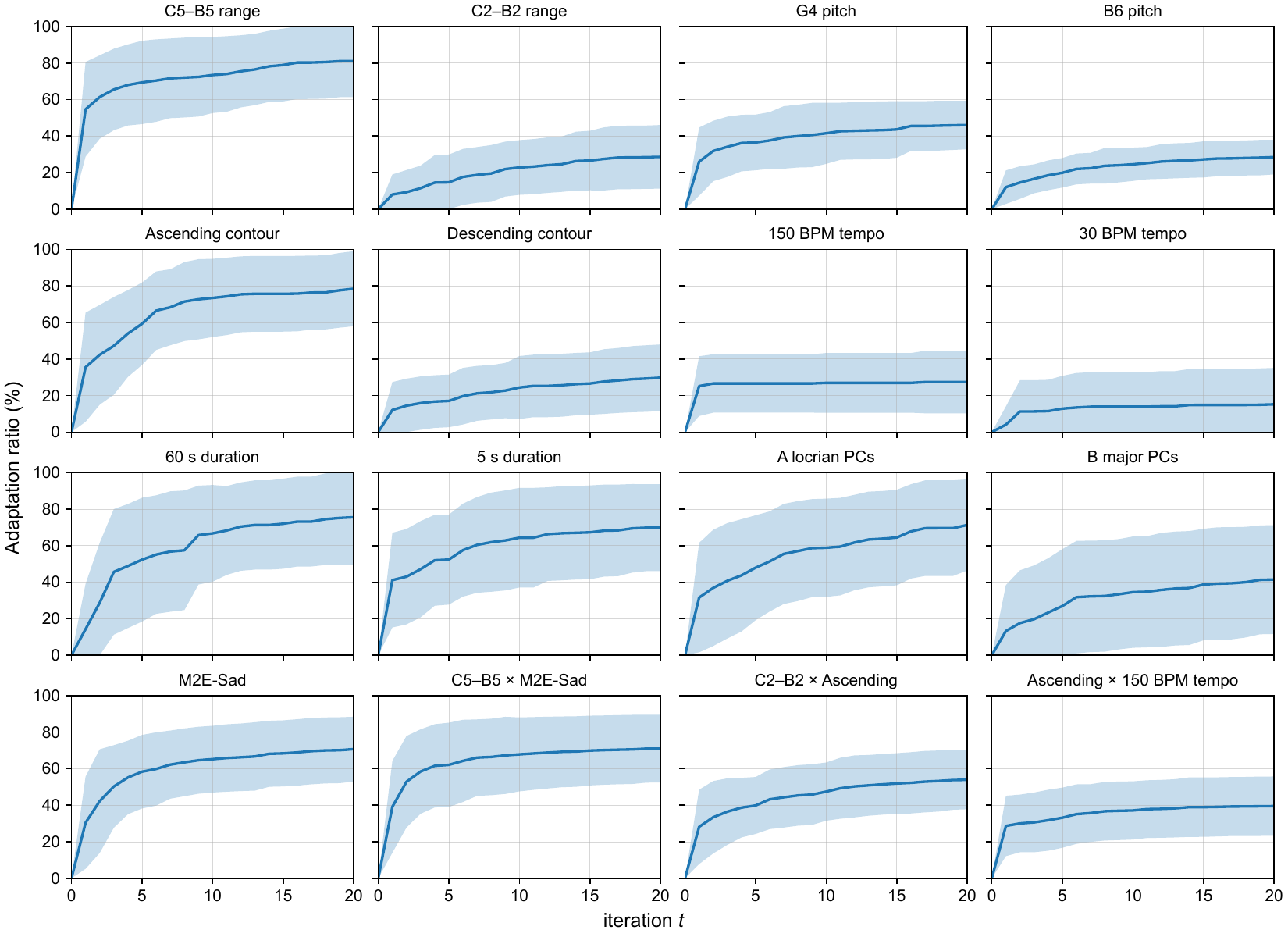}
\caption{Mean adaptation ratio by value function over 20 iterations (30 initial pieces; bands: $\pm 1$ SD).}
\label{fig:exp1_trajectory}
\end{figure*}

Figure~\ref{fig:exp1_trajectory} shows the trajectory of the mean adaptation ratio for each value function. Averaged over all 480 cells, the adaptation ratio was 25.25 at iteration 1, 46.43 at iteration 10, and 51.75 at the final iteration 20 (standard deviation 29.04); gains were large in the early iterations, with gradual improvement continuing in the later half. The trajectories differed across value functions, however: \vflabel{C5--B5 range} and \vflabel{Ascending contour} rose sharply in early iterations, whereas improvement was limited for \vflabel{30~BPM tempo} and \vflabel{150~BPM tempo}.

\begin{figure*}[!t]
\centering
\includegraphics[width=\textwidth]{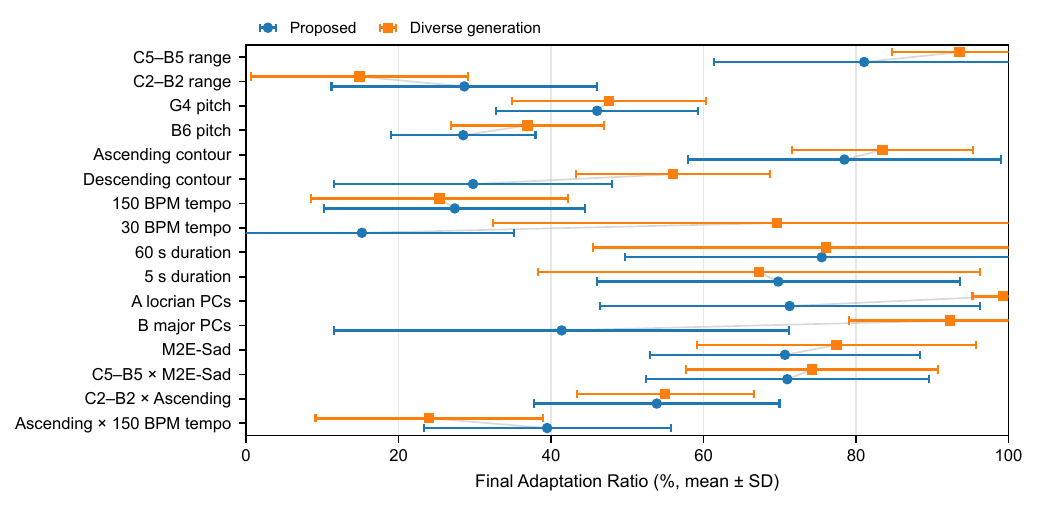}
\caption{Final adaptation ratio by condition and value function (30 initial pieces; error bars: SD; lines connect conditions within functions).}
\label{fig:exp1_condition_comparison}
\end{figure*}

Figure~\ref{fig:exp1_condition_comparison} shows the final adaptation ratios of the proposed and diverse-generation conditions for each value function. In the mixed-effects model pooling all value functions, against the diverse-generation condition's estimated mean of 62.06, the effect of the proposed condition was $\beta=-10.31$ (standard error 1.38, 95\% confidence interval $[-13.02,-7.60]$, $p<0.001$): in the primary analysis averaged over the 16 value functions, the proposed system did not surpass the diverse-generation baseline. In the auxiliary per-function analyses, by contrast, the proposed condition was significantly higher for \vflabel{C2--B2 range} (mean difference in adaptation ratio 13.76, Holm-corrected $p=0.004$) and \vfcomb{Ascending contour}{150~BPM tempo} (mean difference 15.45, $p<0.001$), with no advantage for the remaining 14 functions.

\subsubsection{Additional analyses of adaptation difficulty}
Because the mean adaptation ratio differed substantially across value functions, we analyzed its relation to target atypicality and the initial value gap. Target atypicality is the number of standard deviations by which the value 100 lies from the mean of the value distribution each function assigns to the 1,800 diverse-generation pieces; regarding this distribution as an empirical approximation of what the plain LLM tends to generate under the diversity prompt, larger atypicality means a target state farther from these feedback-free generation tendencies. The initial value gap was defined as $g_{f,m}=100-v_{f,m}^{(0)}$. Target atypicality was standardized across the 16 value functions, the initial value gap within each function across the 30 initial pieces. Over the 480 proposed-condition cells, we estimated a linear mixed-effects model with adaptation ratio as the dependent variable, the two standardized measures as fixed effects, and crossed random intercepts for initial piece and value function.

Table~\ref{tab:exp1_mixedlm} reports the fixed effects and the standard deviations for initial piece, value function, and residual. Target atypicality had $\beta=-11.51$: functions whose targets lie farther from the LLM's generation tendencies adapted less. The initial value gap had $\beta=6.43$: even with the adaptation ratio normalized to the achievable margin, pieces with more room for improvement reached higher ratios within the same function, suggesting that improvement becomes harder near the target. Variation attributable to the initial piece and shared across value functions was relatively small. By contrast, variation across value functions unexplained by target atypicality and residual cell-level variation were both large.

\begin{table*}[!t]
\centering
\small

{
\setlength{\tabcolsep}{6pt}
\renewcommand{\arraystretch}{1.15}

\begin{tabular}[t]{lrrrr}
\multicolumn{5}{c}{Fixed effects} \\
\hline

Effect
& $\beta$
& SE
& 95\% CI
& $p$ \\
\hline

Target atypicality
& $-11.510$
& $4.842$
& $[-21.001,-2.019]$
& $0.017$ \\

Initial value gap
& $6.431$
& $0.869$
& $[4.727,8.134]$
& $<0.001$ \\

\hline
\end{tabular}%
\hspace{2.5em}%
\begin{tabular}[t]{lr}
\multicolumn{2}{c}{Random effects and residual} \\
\hline

Component
& SD \\
\hline

Initial piece
& $2.601$ \\

Value function
& $19.048$ \\

Residual
& $18.632$ \\

\hline
\end{tabular}
}
\caption{Linear mixed-effects model of factors related to adaptation difficulty}
\label{tab:exp1_mixedlm}
\end{table*}

\begin{figure*}[!t]
\centering
\includegraphics[width=\textwidth]{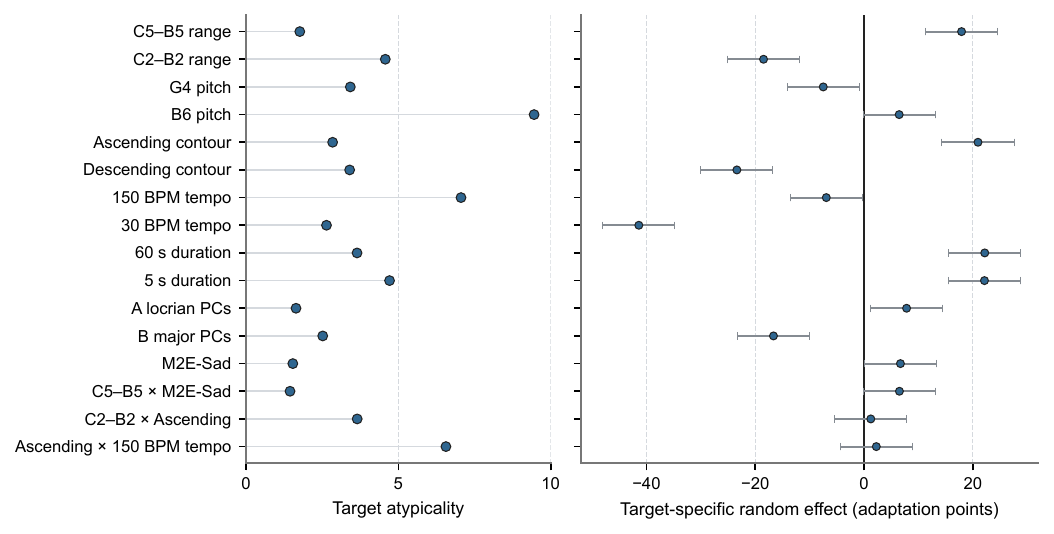}
\caption{Target atypicality (left) and adjusted value-function random effects (right; error bars: conditional 95\% intervals).}
\label{fig:exp1_target_random_effects}
\end{figure*}

Figure~\ref{fig:exp1_target_random_effects} shows target atypicality and each function's conditional random effect. Notably, \vflabel{Ascending contour} and \vflabel{Descending contour}, nearly symmetric in mathematical structure and similar in target atypicality, had random effects of markedly different direction. This asymmetry suggests that adaptation difficulty may depend not only on the value function's mechanical properties, but also on the accessibility of the musical concept and the ease of translating hypotheses into ABC scores. Model-specific tendencies---both in generation and in inferring from the history how musical changes affected rankings---may also have contributed. The random effects are conditional estimates that exploratorily indicate per-function differences unexplained by the fixed effects, not individual tests of between-function differences.

\subsubsection{Ablation of the value-criteria description}
To isolate the contribution of the value-criteria description, we branched all 480 cells at the end of iteration 5 and re-ran iterations 6--10 with the description removed from the prompts, comparing within the same cell against the original run. Because the ranking history is managed as a sliding window, both conditions hold the same number of ranking-history entries over this interval, so any difference reduces to the presence of the description verbalized from that history. For each cell we computed the value growth $v_{f,m}^{(10)}-v_{f,m}^{(5)}$ of the reference piece and tested the condition difference per value function with paired $t$- and Wilcoxon signed-rank tests, Holm-correcting the 16 $p$-values.

The difference averaged only $+0.13$ over all 480 cells (standard deviation 8.17, median 0), and after Holm correction only \vflabel{C2--B2 range} was significant (mean $+0.74$, 95\% confidence interval $[+0.30,+1.19]$). No function was significantly negative, so there was also no evidence that the description hinders adaptation. Thus, while the ranking history remains in context, verbalizing the value criteria neither appreciably accelerated nor impeded adaptation; its information may have been redundant with the history. The description is, however, a compressed representation that can travel beyond the history; Experiment~2 evaluates its contribution in the transfer setting.

\subsubsection{Illustrative case analyses}
To make the adaptation process concrete, we examined four illustrative cases; full iteration-level analyses are provided in Appendix D of the Supplementary Material. Figure~\ref{fig:exp1_ascending_pianoroll} visualizes two \vflabel{Ascending contour} runs. For \texttt{music7}, a hypothesis targeting repetition and dotted rhythms unexpectedly introduced a long ascending sequence, increasing the value from 50.3 to 92.6 in the first iteration. Through subsequent trial and error, the LLM inferred that preserving the specific sequence \texttt{C D E F G A B c...} was important. For \texttt{music10}, simplification from 5 voices to 2 introduced an ascending line, which became fully ascending by iteration 6 and was inherited thereafter. The run reached an adaptation ratio of 100 without the value-criteria descriptions ever identifying ascent. The other cases showed stepwise recognition of a tempo direction while representing a whole-piece target as local changes, and acquisition of cues specific to the Music2Emo evaluator. Thus, higher value arose through qualitatively different paths and did not require identifying the true value function as a general rule.

\begin{figure*}[!t]
\centering
\includegraphics[width=\textwidth]{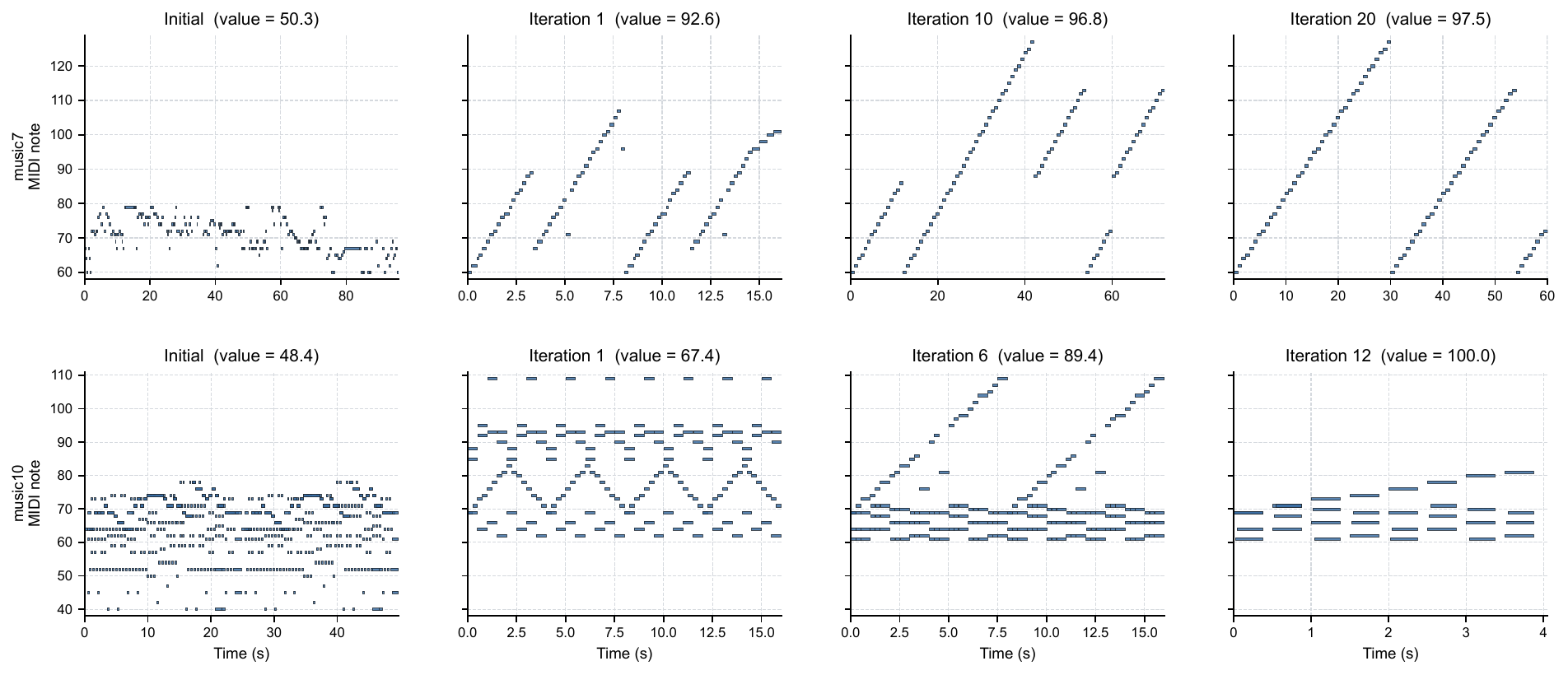}
\caption{Selected reference pieces from two \vflabel{Ascending contour} runs, labeled by generation iteration.}
\label{fig:exp1_ascending_pianoroll}
\end{figure*}

\section{Experiment 2: Transfer of Acquired Value Criteria}\label{sec:exp2}

\subsection{Purpose and setup}
Experiment~2 tests whether the descriptions and histories acquired in Experiment~1 transfer to preference prediction on unseen pieces and generation of new music.

Using the same 30 initial pieces, we selected five value functions spanning adaptation difficulty, target atypicality, and evaluation format (\vflabel{C2--B2 range}, \vflabel{Descending contour}, \vflabel{60~s duration}, \vflabel{M2E-Sad}, \vfcomb{Ascending contour}{150~BPM tempo}). The LLM received no value-function definition or numerical scores, only one of the five contexts in Table~\ref{tab:exp2_conditions}. Conditions (ii)--(iv) separate the description and interaction history acquired in Experiment~1; condition (v) instead supplies ranked examples generated without feedback from the same initial piece.

\begin{table}[!t]
\centering
\small
\begin{tabular}{c>{\raggedright\arraybackslash}p{0.72\columnwidth}}
\hline
Condition & Context given to the LLM \\
\hline
(i) & No rater-preference information \\
(ii) & Final value-criteria description inferred in Experiment~1 \\
(iii) & Full Experiment-1 history: hypotheses, generated pieces, and rankings (20 iterations) \\
(iv) & Final description plus full Experiment-1 history \\
(v) & 60 diverse-generation pieces from the same initial piece, ranked by the simulated rater \\
\hline
\end{tabular}
\caption{Rater-preference information provided to the LLM in Experiment 2.}
\label{tab:exp2_conditions}
\end{table}

\subsection{Prediction task}
\label{subsec:exp2_prediction}

\subsubsection{Task and evaluation method}
The 30 test pieces yielded all $\binom{30}{2}=435$ unordered pairs. For each pair, the LLM selected the piece it predicted the rater would value more highly; we tested both presentation orders. Each of conditions (ii)--(v) comprised 150 cells (5 functions $\times$ 30 initial pieces). Because condition (i) had no function-specific context, its single prediction set was evaluated against all five value functions. For each candidate, we summed its selection log-probabilities across the two presentation orders and chose the piece with the larger total. Pairs tied by either total or value-function score were excluded. For condition (i), we report one accuracy per value function; for conditions (ii)--(v), we average accuracy across the 30 initial-piece cells. Per-function one-sided tests compared condition (iv) with conditions (ii), (iii), and (v) by paired $t$-tests and condition (i) by a one-sample $t$-test, with Holm correction over five functions per comparison. Per-function tests in both tasks were one-sided because the directional hypothesis---that condition (iv), containing both the acquired description and history, outperforms each comparison condition---was specified a priori.

\subsubsection{Accuracy}
Figure~\ref{fig:exp2_prediction_accuracy} shows accuracy per condition (full tables in Appendix C of the Supplementary Material). Mean accuracy stayed near chance (0.5), ranging from 0.46 to 0.62 across conditions and value functions. In the per-function tests, condition (iv) significantly exceeded all four other conditions for \vflabel{C2--B2 range}, conditions (i) and (ii) for \vfcomb{Ascending contour}{150~BPM tempo}, and condition (ii) for \vflabel{60~s duration}, with no significant advantage for \vflabel{Descending contour} or \vflabel{M2E-Sad}. The acquired descriptions and histories thus did not uniformly improve prediction for unseen music; their effect depended on the value function.

\begin{figure*}[!t]
\centering
\includegraphics[width=\textwidth]{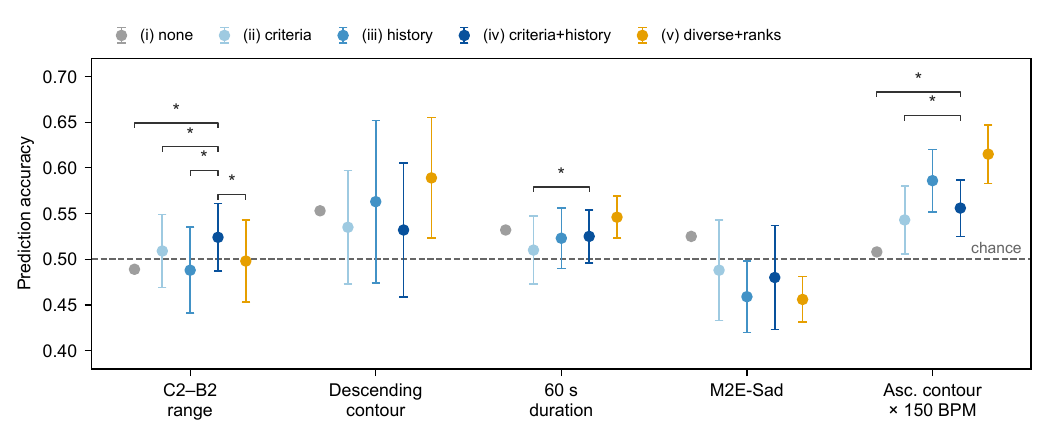}
\caption{Prediction accuracy by condition and value function (error bars: SD; dashed line: chance). Brackets and asterisks indicate pairs in which condition (iv) significantly exceeded the comparison condition (one-sided tests; Holm-adjusted $p<.05$).}
\label{fig:exp2_prediction_accuracy}
\end{figure*}

\subsubsection{Prediction diagnostics}
Three diagnostic analyses further qualified this limited transfer (full results in Appendix C of the Supplementary Material). First, the final adaptation ratio in Experiment~1 was unrelated to condition (iv)'s accuracy after Holm correction, indicating that successful within-loop adaptation did not predict generalization to unseen music. Second, if the LLM used criteria aligned with the value function, accuracy should increase as the value difference $\Delta v=|f(A)-f(B)|$ makes the correct ordering clearer. This expected positive relation appeared weakly for \vflabel{C2--B2 range} and \vfcomb{Ascending contour}{150~BPM tempo}, whereas accuracy decreased with $\Delta v$ for \vflabel{60~s duration} and \vflabel{M2E-Sad}. Because random choices should remain near chance as $\Delta v$ changes, these negative relations suggest that, for more widely separated pairs, the proxy cues used by the LLM became less consistent with the ordering defined by the value function. Third, condition (iv) had the lowest disagreement across presentation orders (25.24\%; 40.92\% in condition (i)), but this greater consistency did not imply greater accuracy. Thus, neither successful within-loop adaptation nor internally consistent choices ensured accurate transfer, and the value-difference analysis further showed that the LLM's proxy cues could diverge from the evaluator's criterion. Direct observability was also insufficient: \vflabel{Descending contour} showed no improvement despite being directly observable in ABC notation.

\subsection{Generation task}
\label{subsec:exp2_generation}

\subsubsection{Task and evaluation method}
Given each context in Table~\ref{tab:exp2_conditions}, the LLM generated music it expected the rater to value highly. Conditions (ii)--(v) produced 30 pieces per cell (18,000 total), each through an independent LLM call with the context fixed and only the sampling seed varied; condition (i) produced a common 30-piece set scored by each function. Exact duplication of a context piece was prohibited, but its musical features could be reused. The cell's 30-piece mean was the primary measure and its maximum the auxiliary measure. One-sided comparisons followed the prediction-task procedure.

\subsubsection{Value of the generated music}
Figure~\ref{fig:exp2b_mean_score} shows the mean value per condition (full tables in Appendix C). For all five functions, condition (iv) was highest, or at the level of the highest, among the four cell-specific-context conditions. It significantly exceeded condition (ii) for all five functions, condition (iii) for the four other than \vflabel{60~s duration}, and condition (i) for the four other than \vflabel{Descending contour}; for \vflabel{60~s duration} it reached 24.39 against 6.62 for condition (i). Against condition (v), however, condition (iv) was significantly higher only for \vflabel{C2--B2 range} and \vfcomb{Ascending contour}{150~BPM tempo}. In contrast to the prediction task, where condition (iv)'s advantage was almost confined to \vflabel{C2--B2 range}, transfer was confirmed for more value functions in generation.

\begin{figure*}[!t]
\centering
\includegraphics[width=\textwidth]{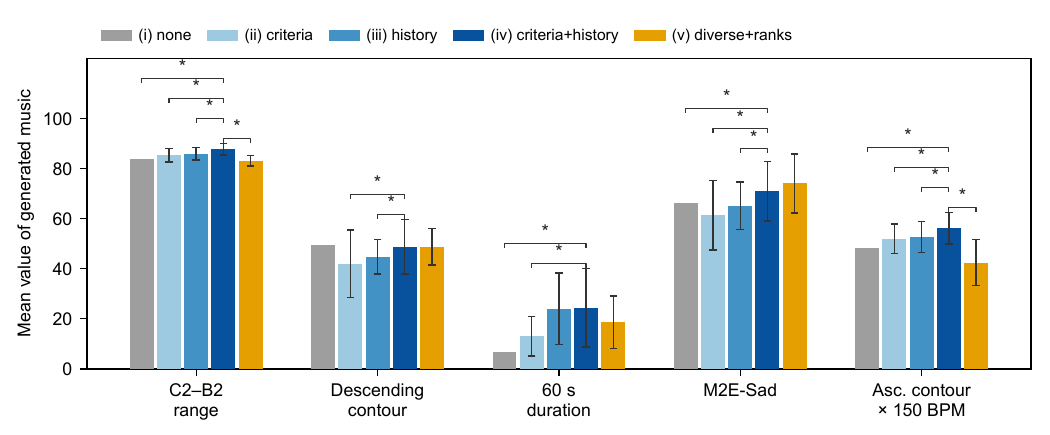}
\caption{Mean generated value by condition and value function (error bars: SD). Brackets and asterisks indicate pairs in which condition (iv) significantly exceeded the comparison condition (one-sided tests; Holm-adjusted $p<.05$).}
\label{fig:exp2b_mean_score}
\end{figure*}

Condition (v) exceeded condition (iv) in mean value for \vflabel{M2E-Sad}, suggesting that ranked examples without explicit verbalization may help for criteria that are difficult to verbalize; however, condition (v)'s poor prediction accuracy prevents a firm interpretation. The acquired context mainly raised average rather than maximum value: context-free condition (i) still produced the highest within-cell maximum for \vflabel{60~s duration} and \vflabel{Descending contour} (Appendix C).

\subsubsection{Acoustic distance to associated reference sets}
To assess whether value gains reflected resemblance to pieces associated with each context rather than criterion inference, we computed, for every generation, the minimum ($d_{\min}$), mean, and maximum cosine distances to an analysis reference set in MERT-v1-95M embedding space \citep{li2024mert}. References were the cell's Experiment-1 generations for conditions (ii)--(iv) and its 60 diverse-generation pieces for condition (v); in condition (ii), the reference pieces were used only for this analysis and were not included in the LLM prompt. History-containing conditions had $d_{\min}$ about 0.03 versus 0.054--0.067 for description-only condition (ii), while mean and maximum distances changed little, suggesting that history drew some generations toward individual examples without shifting the whole distribution. In condition (iv), smaller $d_{\min}$ and larger history-induced reductions were associated with higher values and gains over condition (ii), although \vfcomb{Ascending contour}{150~BPM tempo} improved without this relation (Appendix C). Proximity may therefore have contributed to, but did not fully explain, the gains.

As in the prediction task, the final adaptation ratio in Experiment~1 was only weakly associated with transfer performance: for generated value, only one function--condition comparison remained significant after Holm correction (Appendix C).

\section{Discussion}\label{sec:discussion}

\subsection{Synthesis of the main findings}
Experiment~1 conditionally answers RQ1: adaptation from rankings alone was possible, but surpassed diverse generation only for targets rarely reached without feedback under the diversity prompt. Effectiveness may therefore depend on the relation between the rater's criteria and the LLM's feedback-free generation tendencies.

However, increasing the value of generated music did not imply identifying a generalizable criterion. The case analyses, ablation, and cross-experiment correlations showed that within-loop adaptation did not ensure an accurate description or transfer; the LLM could instead exploit local cues or specific note sequences.

Regarding RQ2, transfer was task-dependent: preference prediction remained near chance, whereas description-plus-history context increased the value of generated music for more functions. The acquired information may therefore be easier to use for generation than general value judgment. Because the loop applies inferred value criteria only through generation, incorporating preference prediction and verification may reduce this asymmetry.

For RQ3, the distance between the target and the LLM's feedback-free generation tendencies, as well as the available improvement margin, was associated with adaptation difficulty. Differences across value functions further suggest that observability in ABC notation and ease of verbalization may affect transfer. Acoustic proximity to referenced pieces was associated with some value gains, but gains also occurred without such proximity. Together, these patterns suggest that difficulty may vary with the relation among the rater, foundation model, and music representation, rather than being an intrinsic property of the value criterion alone. General-purpose LLMs may possess basic musical knowledge while remaining limited in applying it to multi-step reasoning over complete pieces \citep{zhou2024llmsreason}. Thus, although allowing the LLM to form its own representation avoids predefined features, the range of representations it forms may still depend on its ability to recognize, reason about, and verbalize features in symbolic music. The feature-design problem may therefore be displaced to the capabilities of the foundation model rather than eliminated.

\subsection{Strengths and trade-offs of the proposed system}
The principal strength of the system is its ability, under this matched budget of 61 unique evaluated pieces, to guide generation toward targets that simple sampling rarely reaches, while retaining the resulting history and criteria description as reusable context. Because the description can be inspected and corrected by a human, it may also provide an interpretable interface to the adaptation process.

Conversely, generation mediated by hypotheses and language may narrow exploration and bias attention toward local, easily verbalized features. As suggested by the \vflabel{M2E-Sad} result, direct generation from ranked examples may provide a complementary path for criteria that resist verbalization. Combining direction from explicit criteria descriptions with example-driven generation is therefore one promising extension.

\subsection{Limitations}\label{subsec:limitations}
This study has three limitations. First, fixed target-specific functions omit complex, variable human preferences. Because they score only target attributes and place no constraint on overall musical quality, adaptation can converge on musically degenerate solutions; the repetitive and sparse scalar patterns in the \vflabel{Ascending contour} cases illustrate such specification gaming. Human raters may value quality and reject these solutions, motivating human-rater studies, although twenty four-piece rankings may cause fatigue. Feature interactions and context dependence beyond our geometric-mean combinations remain untested. Second, the experiments used one LLM, 60 pieces from one corpus, and five value functions in Experiment~2; generalization to other models, musical styles, and criteria remains unknown. Third, ABC notation constrains criteria involving fine performance expression or acoustic texture.

\subsection{Future work}
The next step is to test the construction and transfer of criteria descriptions with human raters who have complex preferences. Externalized descriptions might also support self-understanding, but their validity should be assessed through both subjective confirmation and objective transfer performance.

The framework may also extend from value to novelty, defined as newness to the rater and operationalized as distance from their listening experience. Because maximizing distance can produce disordered, valueless deviation, desirable direction and degree of deviation should instead form part of the rater's value criteria. Yet deviations far outside the LLM's generative space may remain difficult to infer or realize. Combining value-criteria-based direction with generative-space expansion, as explored by \citet{fujimoto2019generative}, may support music both novel and valuable to the rater.

\section{Conclusion}\label{sec:conclusion}
We proposed a ranking-only iterative in-context learning framework that constructs a rater's value criteria as a natural-language working hypothesis for music generation. Across 480 cells with 16 simulated raters, it exceeded diverse generation only for targets rarely reached by simple sampling. Higher generated value did not imply identifying the criterion as a general rule: acquired context improved generation for unseen music, while preference prediction remained near chance. The effectiveness and limits of ranking-only adaptation therefore depend on the foundation model's ability to recognize and verbalize musical features.

\section{Reproducibility}
Because the simulated raters are fully specified value functions, evaluation is deterministic. All LLM calls used gpt-oss-120b; Appendix E reports the decoding, seed, model-revision, and software settings. The 60 OpenScore Lieder Corpus pieces came from the GitHub mirror at commit \texttt{6b2dc542ce2e}; audio used FluidSynth and the FluidR3 GM SoundFont (v3.1-5.3, SHA256 \texttt{74594e8f\allowbreak 4250680a\allowbreak df590507\allowbreak a306655a\allowbreak 29993534\allowbreak 3583256f\allowbreak 3b722c48\allowbreak a1bc1cb0}). Because inference may vary with concurrent batch composition even under fixed seeds, individual runs may not reproduce exactly. Code and generated data will be deposited in an open repository with a DOI; the repository link will be added in a future revision of this preprint.

\section*{Artificial Intelligence Use}
Generative AI tools were used for translation, language editing, and an initial visual draft of Figure~\ref{fig:system_overview}, which the authors subsequently reconstructed and manually revised. The authors reviewed and approved all scientific claims, analyses, citations, final text, and figures. No research data were generated or altered using these tools.

\section*{Competing Interests}
The authors have no competing interests to declare.

\bibliographystyle{plainnat}
\bibliography{references}

\clearpage
\section*{Supplementary Material}
\def\ARXIVCOMBINED{}
\ifdefined\ARXIVCOMBINED
\else
\documentclass[11pt]{article}

\usepackage[margin=25mm]{geometry}

\usepackage{iftex}
\ifLuaTeX
  \usepackage{luatexja}
  \usepackage[haranoaji]{luatexja-preset}
\else
  \errmessage{Compile this file with LuaLaTeX.}
\fi

\usepackage{amsmath,amsfonts}
\usepackage{url}
\usepackage{graphicx}
\usepackage{array}
\usepackage{listings}
\emergencystretch=2em
\fi

\ifdefined\ARXIVCOMBINED
  \newcommand{\finishsupplement}{ }
\else
  \newcommand{\finishsupplement}{\end{document}}
\fi

\lstset{
  basicstyle=\ttfamily\scriptsize,
  breaklines=true,
  breakindent=0pt,
  columns=fullflexible,
  keepspaces=true,
  upquote=true,
  frame=single,
  framesep=4pt,
  aboveskip=8pt,
  belowskip=8pt,
  literate={♭}{{$\flat$}}1 {　}{{ }}1,
}

\ifdefined\ARXIVCOMBINED
\else
\newcommand{\vflabel}[1]{\textit{#1}}
\newcommand{\vfcomb}[2]{\vflabel{#1}~\ensuremath{\times}~\vflabel{#2}}
\fi

\newcounter{prompt}
\renewcommand{\theprompt}{S\arabic{prompt}}
\newcommand{\promptheading}[2]{%
  \refstepcounter{prompt}\label{#1}%
  \par\medskip\noindent\textbf{Prompt~\theprompt:}~#2\par\nobreak\vspace{2pt}%
}

\renewcommand{\thesection}{\Alph{section}}
\newcommand{\suppappendix}[1]{\section*{Appendix \stepcounter{section}\thesection : #1}}
\setcounter{section}{0}

\ifdefined\ARXIVCOMBINED
\else
\title{Supplementary Material\\
{\large Inferring Value Criteria from Ordinal Preferences:\\
An Iterative In-Context Learning Framework for Music Generation}}
\author{}
\date{}

\begin{document}

\maketitle
\thispagestyle{empty}
\fi

\suppappendix{Full Prompts}

This appendix provides the full text of the prompts that the proposed system and the diverse-generation baseline of Experiment~1 give to the LLM. Prompt~\ref{prompt:system} is the system prompt shared by all LLM calls of the proposed system; it establishes the composer role and the rules for writing ABC notation. The hypothesis-generation step uses Prompt~\ref{prompt:initial_hypothesis} in the first iteration and Prompt~\ref{prompt:hypothesis} from the second iteration onward. The music-generation step uses Prompt~\ref{prompt:generate_abc}, and the value-criteria inference step uses Prompt~\ref{prompt:infer}. The diverse-generation baseline uses Prompt~\ref{prompt:diversity_system} as the system prompt and Prompt~\ref{prompt:diversity_generation} for music generation.

Prompts are reproduced verbatim as used in the experiments; the gap in rule numbering (10 to 12) reflects the prompts actually given to the LLM.

In the prompts, \texttt{\{\{...\}\}} denotes placeholders replaced at run time. \texttt{\{\{initial\_abc\}\}} is replaced with the ABC score of the initial piece, \texttt{\{\{history\}\}} with the ranking history, \texttt{\{\{iteration\}\}} with the current iteration number, and \texttt{\{\{hypothesis\}\}} with the single hypothesis to compose from. \texttt{\{\{value\_criteria\}\}} receives the latest value-criteria description, preceded by a preamble stating that it is a tentative working hypothesis derived from trial and error (it is an empty string while no description has been generated yet). In Prompt~\ref{prompt:diversity_generation}, \texttt{\{\%...\%\}} is control syntax for conditionals and loops; it expands the sequence of ABC scores of the most recent generations (\texttt{previous\_abcs}) one piece at a time. \texttt{\{\{num\_previous\}\}} is replaced with the number of generated pieces included in the context, and \texttt{\{\{generation\_index\}\}} with the current generation number.

\promptheading{prompt:system}{System prompt (shared by all LLM calls)}
\begin{lstlisting}
You are a composer who creates music using abc notation. Your objective is to understand the evaluator's musical preferences and the value criteria underlying the evaluations. To do so, you can repeat the process of composing multiple pieces of music and having the evaluator rank them. You must infer why certain music is preferred over others to understand such criteria, and become capable of generating new music that the evaluator finds more valuable.

To achieve this objective, please repeat the following process:
1. Hypothesize:
   Based on the evaluator's history of evaluating the music you have generated so far, formulate hypotheses about what kind of music would allow you to understand the evaluator's value criteria, and what the evaluator is likely to find more valuable.
2. Compose new pieces:
   Compose new music in ABC notation based on the hypotheses you formulated.

For the composition of abc notation, please follow the rules below.
1.  Header information:
    -   Start with X: music identifier.
    -   Include the minimum required elements:
        -   T: title
        -   M: meter
        -   L: default note length
        -   Q: tempo
        -   K: key
    -   Using X: music identifier and T: title more than once or reusing them for different parts of the piece is prohibited.
    -   Ensure that the key part of the K: field (e.g., D, G) is written in uppercase. Writing the key part of the K: field in lowercase is prohibited.
    -   Prefer putting key, meter, default-length, and tempo changes on their own lines.
    -   Inline key or clef directives such as [K:G] or [K:D] may appear in reference ABC files and may be used when musically necessary, but avoid excessive inline changes.
2.  Defining voices:
    -   If you define voices, use V:.
    -   When defining multiple voices, increment the numbers to clarify (e.g., V:1 clef=treble, V:2, V:3, etc.).
    -   Simply adding a text label after V: is insufficient; be sure to define each voice with a number.
    -   When multiple voices (V:) are played simultaneously, ensure that the repeat structure is aligned across all voices.
    -   Skipping voice numbers when defining voices (e.g., defining V:1, V:2, then V:4) is prohibited.
    -   When using multiple voices (V:1, V:2, ...), all bar markers (|, |:, :|) must appear inside V:N music lines. Standalone bar lines outside any V:N content (for example, a line containing only `|:` or `| % comment`) confuse abc2midi and cause most of the music to be silently dropped.
3.  Chord representation:
    -   You may consider chord progressions, but using chord symbols (e.g., "C") is prohibited.
    -   Use square brackets [] to represent notes played simultaneously (e.g., C major chord C, E, G is [C E G]).
    -   Square brackets may also appear in valid inline ABC directives such as [K:G].
    -   Alternatively, you can separate voices to express melody and accompaniment.
    -   Using chord notation [] for single notes is prohibited.
        -   Correct: [C E G] (chord)
    -   Using broken rhythm (>) or grace notes ({}) inside chord notation [] is prohibited.
    -   Applying rolls or trills to chords is prohibited.
    -   Nesting square brackets [] is prohibited.
4.  Accidentals:
    -   If you need to raise or lower a note by a semitone, use ^ for a sharp and _ for a flat.
    -   Place accidentals (^, _, =) immediately before the note without any space.
        -   Correct: ^f (sharp)
    -   In the K: field, you should use # and ♭ for sharps and flats.
    -   Attaching both sharps and flats to the same note is prohibited.
5.  Meter consistency:
    -   Ensure that the sum of the note lengths in each measure matches the meter.
6.  Tempo
    -   Write the tempo clearly by specifying the beat unit.
    -   Writing the tempo using tempo markings is prohibited.
7.  Repeats and Multi-Endings:
    -   Use |: and :| accurately. Each |: must be correctly closed with a corresponding :|. Omitting the opening |: is prohibited.
    -   Nested repeats are prohibited.
    -   When writing first and second endings, use:
        -   |1 ... :| for the first ending and repeat.
        -   |2 ... || for the second ending (or :| if repeating again).
        -   Inserting spaces between | and the number (e.g., | 1, | 2) is prohibited.
            -   Correct: |1, |2
    -   Using [] to indicate multi-endings is prohibited.
    -   Repeat markers (|:, :|) cannot share a line with header directives (K:, M:, L:, Q:). If you need to change key/meter at a repeat point, use two separate lines: the music line with |:, and the directive on its own line.
    -   A bar marker (|, |:, :|) on a line by itself, or with only a `% comment`, is prohibited. Bar markers must always accompany music content on the same line.
8.  Octave adjustment methods:
    -   To raise an octave, change the basic note letter to lowercase (e.g., changing C to c raises from C4 to C5).
    -   To raise further octaves, add an apostrophe (') after the lowercase letter.
    -   To lower an octave, add a comma (,) after the note (e.g., changing C to C, lowers from C4 to C3).
    -   You can add multiple apostrophes or commas to raise or lower additional octaves.
    -   Place octave modification symbols before the note length.
        -   Correct: B,2
    -   Octave modifications should only go in one direction. Using both apostrophes (') and commas (,) on the same note for octave modification **is prohibited**.
9. MIDI configuration:
    -   Using non-standard commands like %%MIDI tempo 100 to specify tempo is prohibited. The Q: field must be used.
    -   MIDI instrument numbers may be used if necessary. The MIDI program change numbers range from 0 to 127.
    -   To specify an instrument, write the following after K: key (e.g., for a violin):
        `%%MIDI program 40`
    -   To apply an instrument to a specific voice (e.g., V1), use the following format:
        `V:1`
        `%%MIDI program 40`
    -   Violin is just an example; other instruments can be specified using appropriate MIDI program numbers.
10. Other Important Rules
    -   The only allowed dynamic markings are !pp!, !p!, !mp!, !mf!, !f!, and !ff!. abc2midi reflects these standard dynamics in MIDI velocity.
    -   The only allowed non-dynamic decorations are MIDI-relevant decorations used in this workflow: !<(!, !<)!, !>(!, !>)!, !fermata!, !arpeggio!, !trill!, !ped!, and !ped-up!.
    -   Do not invent unsupported !...! decorations.
    -   Use z for ordinary rests. Attaching octave symbols to rests is prohibited.　Do not use x as a note pitch.
    -   When using dotted notes, ensure the total beats per measure are correct, as exceeding the measure's beat count may cause the dot to be ignored.
    -   When specifying note length as a fraction, only use denominators that are powers of 2 (e.g., 2, 4, 8).
        -   Correct: e/2 A/4
    -   Using the slash (/) for anything other than specifying note length, key (K:), or tempo (Q:) is prohibited.
    -   When tying notes, connect notes of the same pitch with a hyphen -.
        -   Correct: e2- e2
    -   When using slurs, enclose notes in parentheses.
        -   Correct: (c d e)
    -   Apply broken rhythm (>) only between notes of the same length. Inserting the broken rhythm symbol (>) between the note and its duration is prohibited.
        -   Correct: e>f
    -   Using x or i as note names is prohibited.
    -   Nested tuplets are prohibited.
    -   Including rests (z), fractional note lengths (e.g., e/2, f3/2), or articulation marks within tuplets is prohibited. 
        -   Correct: (3ABC
    -   Tuplets must consist entirely of notes of the same duration.
    -   Inserting a space between the note and its duration is prohibited.
    -   If you wish to indicate parts of the music, use P: (not mandatory).
    -   Do not write prose explanations inside the ABC.
12. Content of abc notation:
    -   When writing ABC notation, include only ABC-related content.
    -   Write only one piece.
    -   Adding non-ABC content within the notation is prohibited.
    -   Comments should be strictly limited to MIDI-related information.
\end{lstlisting}

\promptheading{prompt:initial_hypothesis}{Hypothesis generation (first iteration)}
\begin{lstlisting}
The following music is provided to you as the initial material for comparison:
{{ initial_abc }}

------
This is iteration {{ iteration }}.
Formulate three hypotheses regarding what kind of music would allow you to understand the evaluator's value criteria, and what the evaluator is likely to find more valuable. The hypotheses formulated here will be used to compose new pieces in the next step, creating a separate piece of music for each individual hypothesis. Since the actual composition will be done in a later stage, please output only the hypotheses here.

Please note the following:
- Please formulate your hypotheses specifically as modifications to the initial music. Describe how changing that specific piece would lead to a higher evaluation or help you better understand the evaluator's value criteria. The scope of these modifications can range from minor adjustments to fundamental changes that alter the entire structure.
- The description of the hypotheses can be either abstract or concrete, as long as you are able to compose music in ABC notation based on them.
- Generating music that is identical to what you have previously composed is prohibited. As long as they are not exactly the same, the newly composed music may be either similar to or completely different from your past compositions.

Based on the information above, please write your Hypotheses in the following format:
Hypothesis 1:
Hypothesis 2:
Hypothesis 3:
\end{lstlisting}

\promptheading{prompt:hypothesis}{Hypothesis generation (second and later iterations)}
\begin{lstlisting}
{{ value_criteria }}

Below is a sequential history of the evaluator's rankings of the music you composed based on your hypotheses. For comparison, each ranking includes the music that was most preferred at that time (the one that was top-ranked in the previous iteration).
{{ history }}

------
This is iteration {{ iteration }}.
Formulate three hypotheses regarding what kind of music would allow you to understand the evaluator's value criteria, and what the evaluator is likely to find more valuable. The hypotheses formulated here will be used to compose new pieces in the next step, creating a separate piece of music for each individual hypothesis. Since the actual composition will be done in a later stage, please output only the hypotheses here.

Please note the following:
- Please formulate your hypotheses specifically as modifications to the music that was ranked highest in the previous iteration. Describe how changing that specific piece would lead to a higher evaluation or help you better understand the evaluator's value criteria. The scope of these modifications can range from minor adjustments to fundamental changes that alter the entire structure.
- The description of the hypotheses can be either abstract or concrete, as long as you are able to compose music in ABC notation based on them.
- Generating music that is identical to what you have previously composed is prohibited. As long as they are not exactly the same, the newly composed music may be either similar to or completely different from your past compositions.

Based on the information above, please write your Hypotheses in the following format:
Hypothesis 1:
Hypothesis 2:
Hypothesis 3:
\end{lstlisting}

\promptheading{prompt:generate_abc}{Music generation}
\begin{lstlisting}
{{ value_criteria }}

Below is a sequential history of the evaluator's rankings of the music you composed based on your hypotheses. For comparison, each ranking includes the music that was most preferred at that time (the one that was top-ranked in the previous iteration).
{{ history }}

------
This is iteration {{ iteration }}.

One of your hypotheses regarding what kind of music would allow you to understand the evaluator's value criteria, and what the evaluator is likely to find more valuable:
{{ hypothesis }}

Using the above information, please compose a piece using ABC notation. Follow the rules of ABC notation and generate only the content of the ABC score.
\end{lstlisting}

\promptheading{prompt:infer}{Value-criteria inference}
\begin{lstlisting}
{{ value_criteria }}

Below is a sequential history of the evaluator's rankings of the music you composed based on your hypotheses. For comparison, each ranking includes the music that was most preferred at that time (the one that was top-ranked in the previous iteration).
{{ history }}

------
This is iteration {{ iteration }}.

Based on the information above, infer and organize the value criteria underlying the evaluator's rankings. When inferring, pay attention not only to the hypotheses you formed at each iteration but also to the composed ABC notation itself. This description will be referenced when formulating hypotheses and composing music in subsequent iterations.
\end{lstlisting}

\promptheading{prompt:diversity_system}{System prompt (diverse-generation baseline)}
\begin{lstlisting}
You are a composer who creates varied pieces of music using abc notation. Each time you compose a new piece, your goal is simply to produce music that is meaningfully different from the reference piece and from any pieces you have already composed in this session.

For the composition of abc notation, please follow the rules below.
1.  Header information:
    -   Start with X: music identifier.
    -   Include the minimum required elements:
        -   T: title
        -   M: meter
        -   L: default note length
        -   Q: tempo
        -   K: key
    -   Using X: music identifier and T: title more than once or reusing them for different parts of the piece is prohibited.
    -   Ensure that the key part of the K: field (e.g., D, G) is written in uppercase. Writing the key part of the K: field in lowercase is prohibited.
    -   Prefer putting key, meter, default-length, and tempo changes on their own lines.
    -   Inline key or clef directives such as [K:G] or [K:D] may appear in reference ABC files and may be used when musically necessary, but avoid excessive inline changes.
2.  Defining voices:
    -   If you define voices, use V:.
    -   When defining multiple voices, increment the numbers to clarify (e.g., V:1 clef=treble, V:2, V:3, etc.).
    -   Simply adding a text label after V: is insufficient; be sure to define each voice with a number.
    -   When multiple voices (V:) are played simultaneously, ensure that the repeat structure is aligned across all voices.
    -   Skipping voice numbers when defining voices (e.g., defining V:1, V:2, then V:4) is prohibited.
    -   When using multiple voices (V:1, V:2, ...), all bar markers (|, |:, :|) must appear inside V:N music lines. Standalone bar lines outside any V:N content (for example, a line containing only `|:` or `| % comment`) confuse abc2midi and cause most of the music to be silently dropped.
3.  Chord representation:
    -   You may consider chord progressions, but using chord symbols (e.g., "C") is prohibited.
    -   Use square brackets [] to represent notes played simultaneously (e.g., C major chord C, E, G is [C E G]).
    -   Square brackets may also appear in valid inline ABC directives such as [K:G].
    -   Alternatively, you can separate voices to express melody and accompaniment.
    -   Using chord notation [] for single notes is prohibited.
        -   Correct: [C E G] (chord)
    -   Using broken rhythm (>) or grace notes ({}) inside chord notation [] is prohibited.
    -   Applying rolls or trills to chords is prohibited.
    -   Nesting square brackets [] is prohibited.
4.  Accidentals:
    -   If you need to raise or lower a note by a semitone, use ^ for a sharp and _ for a flat.
    -   Place accidentals (^, _, =) immediately before the note without any space.
        -   Correct: ^f (sharp)
    -   In the K: field, you should use # and ♭ for sharps and flats.
    -   Attaching both sharps and flats to the same note is prohibited.
5.  Meter consistency:
    -   Ensure that the sum of the note lengths in each measure matches the meter.
6.  Tempo
    -   Write the tempo clearly by specifying the beat unit.
    -   Writing the tempo using tempo markings is prohibited.
7.  Repeats and Multi-Endings:
    -   Use |: and :| accurately. Each |: must be correctly closed with a corresponding :|. Omitting the opening |: is prohibited.
    -   Nested repeats are prohibited.
    -   When writing first and second endings, use:
        -   |1 ... :| for the first ending and repeat.
        -   |2 ... || for the second ending (or :| if repeating again).
        -   Inserting spaces between | and the number (e.g., | 1, | 2) is prohibited.
            -   Correct: |1, |2
    -   Using [] to indicate multi-endings is prohibited.
    -   Repeat markers (|:, :|) cannot share a line with header directives (K:, M:, L:, Q:). If you need to change key/meter at a repeat point, use two separate lines: the music line with |:, and the directive on its own line.
    -   A bar marker (|, |:, :|) on a line by itself, or with only a `% comment`, is prohibited. Bar markers must always accompany music content on the same line.
8.  Octave adjustment methods:
    -   To raise an octave, change the basic note letter to lowercase (e.g., changing C to c raises from C4 to C5).
    -   To raise further octaves, add an apostrophe (') after the lowercase letter.
    -   To lower an octave, add a comma (,) after the note (e.g., changing C to C, lowers from C4 to C3).
    -   You can add multiple apostrophes or commas to raise or lower additional octaves.
    -   Place octave modification symbols before the note length.
        -   Correct: B,2
    -   Octave modifications should only go in one direction. Using both apostrophes (') and commas (,) on the same note for octave modification **is prohibited**.
9. MIDI configuration:
    -   Using non-standard commands like %%MIDI tempo 100 to specify tempo is prohibited. The Q: field must be used.
    -   MIDI instrument numbers may be used if necessary. The MIDI program change numbers range from 0 to 127.
    -   To specify an instrument, write the following after K: key (e.g., for a violin):
        `%%MIDI program 40`
    -   To apply an instrument to a specific voice (e.g., V1), use the following format:
        `V:1`
        `%%MIDI program 40`
    -   Violin is just an example; other instruments can be specified using appropriate MIDI program numbers.
10. Other Important Rules
    -   The only allowed dynamic markings are !pp!, !p!, !mp!, !mf!, !f!, and !ff!. abc2midi reflects these standard dynamics in MIDI velocity.
    -   The only allowed non-dynamic decorations are MIDI-relevant decorations used in this workflow: !<(!, !<)!, !>(!, !>)!, !fermata!, !arpeggio!, !trill!, !ped!, and !ped-up!.
    -   Do not invent unsupported !...! decorations.
    -   Use z for ordinary rests. Attaching octave symbols to rests is prohibited.　Do not use x as a note pitch.
    -   Do not use layout-only information fields such as I:linebreak, [I:staff -1], or [I:staff +1]. They do not change the MIDI sound in this workflow.
    -   When using dotted notes, ensure the total beats per measure are correct, as exceeding the measure's beat count may cause the dot to be ignored.
    -   When specifying note length as a fraction, only use denominators that are powers of 2 (e.g., 2, 4, 8).
        -   Correct: e/2 A/4
    -   Using the slash (/) for anything other than specifying note length, key (K:), or tempo (Q:) is prohibited.
    -   When tying notes, connect notes of the same pitch with a hyphen -.
        -   Correct: e2- e2
    -   When using slurs, enclose notes in parentheses.
        -   Correct: (c d e)
    -   Apply broken rhythm (>) only between notes of the same length. Inserting the broken rhythm symbol (>) between the note and its duration is prohibited.
        -   Correct: e>f
    -   Using x or i as note names is prohibited.
    -   Nested tuplets are prohibited.
    -   Including rests (z), fractional note lengths (e.g., e/2, f3/2), or articulation marks within tuplets is prohibited.
        -   Correct: (3ABC
    -   Tuplets must consist entirely of notes of the same duration.
    -   Inserting a space between the note and its duration is prohibited.
    -   If you wish to indicate parts of the music, use P: (not mandatory).
    -   Do not write prose explanations inside the ABC.
12. Content of abc notation:
    -   When writing ABC notation, include only ABC-related content.
    -   Write only one piece.
    -   Adding non-ABC content within the notation is prohibited.
    -   Comments should be strictly limited to MIDI-related information.
\end{lstlisting}

\promptheading{prompt:diversity_generation}{Music generation (diverse-generation baseline)}
\begin{lstlisting}
Below is the abc notation you will use as a reference (the initial piece):

{{ initial_abc }}

------
{% if previous_abcs %}
You have already composed the following {{ num_previous }} pieces in this session:

{% for prev in previous_abcs %}
=== Previously composed piece {{ loop.index }} ===
{{ prev }}

{% endfor %}
------
{% endif %}
This is generation {{ generation_index }}.

Compose ONE new piece in ABC notation that is meaningfully different from the reference piece above{% if previous_abcs %} and from every previously composed piece you have already produced in this session{% endif %}. Aim for genuine variety: avoid creating pieces that are nearly identical to the reference or to your earlier compositions.

Follow the rules of ABC notation. Generate only the content of one ABC score (no commentary, no hypothesis, no analysis).
\end{lstlisting}

\suppappendix{Processing Details of the Value Functions}

This appendix details the processing of each of the 13 single value functions introduced in Section~4 of the main text. Each value function takes as input the ABC score of the music under evaluation and the MIDI file converted from it with abc2midi, and returns a value from 0 to 100. Except for the tempo value functions (which read their feature from the ABC header) and \vflabel{M2E-Sad} (which computes its value from the acoustic signal), the 10 functions extract their features from the MIDI file. Tempo is expressed in beats per minute (BPM). MIDI feature extraction collects the note-on events of all tracks (velocity $>0$, excluding percussion channels), and we write the resulting sequence of MIDI note numbers as $p_1, \dots, p_N$. If the ABC-to-MIDI conversion fails, or the music contains no notes, the value is 0.

The value functions compute their values in one of two ways. The \emph{distance type} (the 8 pitch-range, pitch, tempo, and duration functions) computes a distance $d$ between the feature and the target state and maps it to 0--100 by exponential decay
\begin{equation}
v = 100 \exp(-\lambda d)
\label{eq:decay}
\end{equation}
with decay rate $\lambda = 0.1$ by default and $\lambda = 0.025$ for tempo only (justified below). The \emph{ratio type} (the 4 pitch-contour and pitch-class functions) uses the proportion (0--1) of elements satisfying the target state, multiplied by 100. The remaining \vflabel{M2E-Sad} scales the output probability of a pretrained model by 100. The details follow, by category.

\subsection*{B.1 Pitch range (\vflabel{C5--B5 range}, \vflabel{C2--B2 range})}

The target state is that all pitches lie within a specific one-octave interval. The target interval $[l, u]$ is $[72, 83]$ in MIDI note numbers for \vflabel{C5--B5 range} and $[36, 47]$ for \vflabel{C2--B2 range}. For each note $p_i$, the distance $d_i$ from the target interval is defined in octaves as
\begin{equation}
d_i =
\begin{cases}
0 & (l \le p_i \le u) \\
(l - p_i)/12 & (p_i < l) \\
(p_i - u)/12 & (p_i > u)
\end{cases}
\end{equation}
and the value is the per-note mean of Eq.~\eqref{eq:decay} ($\lambda = 0.1$):
\begin{equation}
v = \frac{100}{N} \sum_{i=1}^{N} \exp(-0.1\, d_i).
\end{equation}
The value is 100 when all notes lie within the interval.

\begin{center}
\begin{minipage}[t]{0.42\linewidth}\centering
\includegraphics[width=\linewidth]{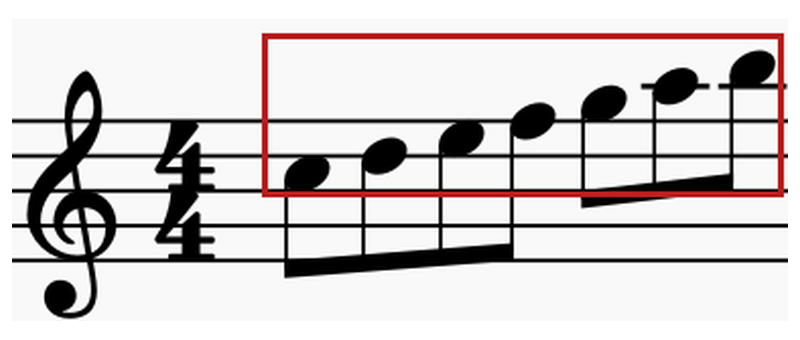}\\
{\small \vflabel{C5--B5 range}}
\end{minipage}\hspace{0.06\linewidth}
\begin{minipage}[t]{0.42\linewidth}\centering
\includegraphics[width=\linewidth]{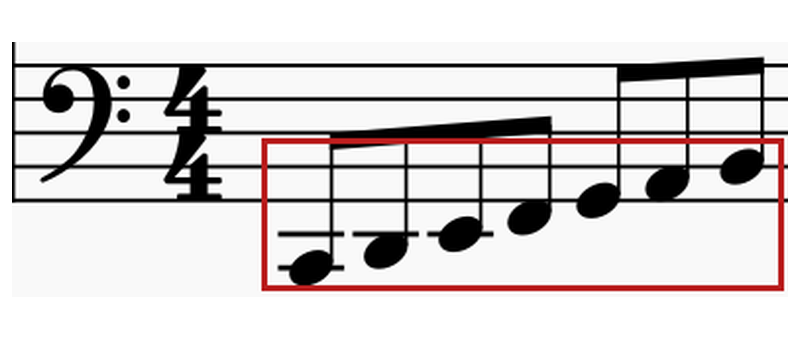}\\
{\small \vflabel{C2--B2 range}}
\end{minipage}
\end{center}

\subsection*{B.2 Pitch (\vflabel{G4 pitch}, \vflabel{B6 pitch})}

The target state is that all pitches equal a specific single pitch. The target pitch $n$ is MIDI note number 67 for \vflabel{G4 pitch} and 95 for \vflabel{B6 pitch}. Defining the distance of each note $p_i$ to the target in semitones as $d_i = |p_i - n|$, the value is, as for pitch range, the per-note mean of the exponential decay ($\lambda = 0.1$):
\begin{equation}
v = \frac{100}{N} \sum_{i=1}^{N} \exp(-0.1\, |p_i - n|).
\end{equation}

\begin{center}
\begin{minipage}[t]{0.42\linewidth}\centering
\includegraphics[width=\linewidth]{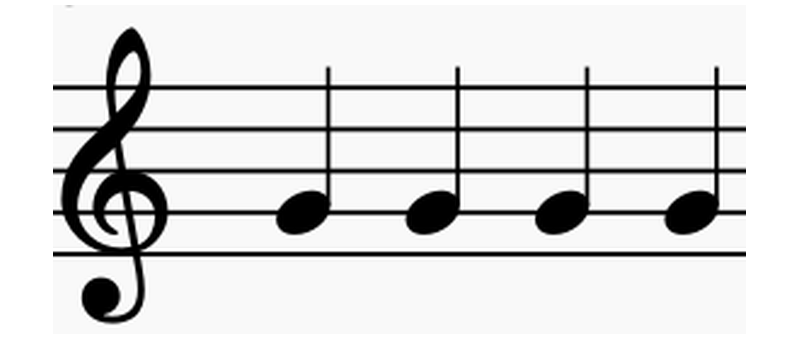}\\
{\small \vflabel{G4 pitch}}
\end{minipage}\hspace{0.06\linewidth}
\begin{minipage}[t]{0.42\linewidth}\centering
\includegraphics[width=\linewidth]{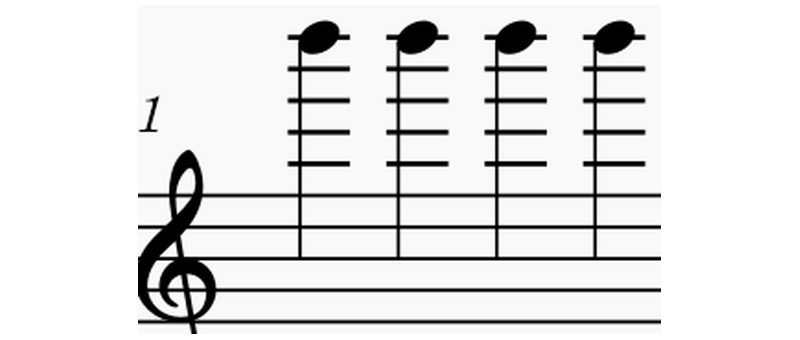}\\
{\small \vflabel{B6 pitch}}
\end{minipage}
\end{center}

\subsection*{B.3 Pitch contour (\vflabel{Ascending contour}, \vflabel{Descending contour})}

The melodic line is first extracted from the MIDI file as the sequence of transitions of the highest note sounding at each moment. The note-on and note-off events of all tracks are merged in order of absolute time, the set of currently sounding notes is tracked at each moment, and the highest note is recorded only when it changes from the previous one (all events at the same time are applied before taking the highest note, so the result does not depend on the key order within a chord). Writing the resulting sequence of highest notes as $m_1, \dots, m_K$, we count the ascending transitions $A = |\{j \mid m_{j+1} > m_j\}|$ and the descending transitions $D = |\{j \mid m_{j+1} < m_j\}|$. The value is the proportion of ascending (descending) transitions among all transitions:
\begin{equation}
v = 100 \cdot \frac{A}{A + D} \quad \text{($100 \cdot \tfrac{D}{A+D}$ for \vflabel{Descending contour})}.
\end{equation}
If $K < 2$, or no transitions exist, $v = 0$.

\begin{center}
\begin{minipage}[t]{0.42\linewidth}\centering
\includegraphics[width=\linewidth]{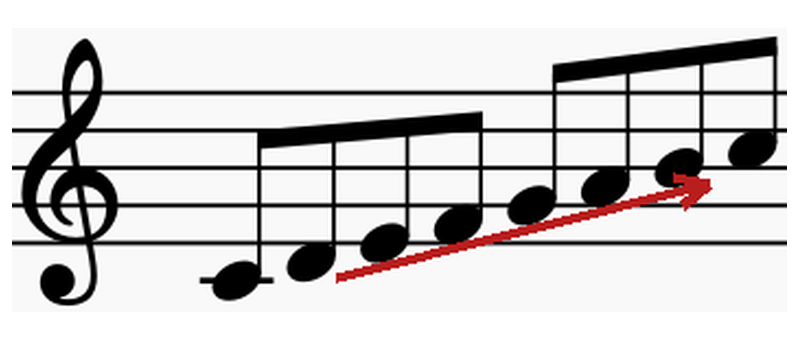}\\
{\small \vflabel{Ascending contour}}
\end{minipage}\hspace{0.06\linewidth}
\begin{minipage}[t]{0.42\linewidth}\centering
\includegraphics[width=\linewidth]{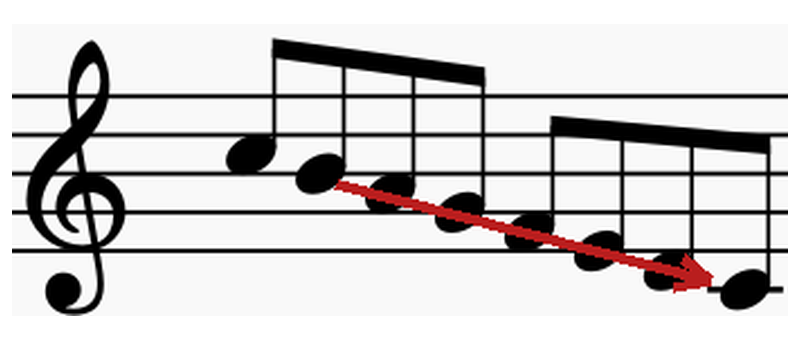}\\
{\small \vflabel{Descending contour}}
\end{minipage}
\end{center}

\subsection*{B.4 Tempo (\vflabel{150~BPM tempo}, \vflabel{30~BPM tempo})}

Tempo is read from the \texttt{Q} field of the ABC header. The form \texttt{Q:1/4=120} is interpreted as BPM with the specified note value converted to quarter notes, and the form \texttt{Q:120} as quarter-note BPM. If several \texttt{Q} fields are present, their mean is used; if none is present, the ABC-standard default of 120~BPM is taken as the piece's tempo $T$. Eq.~\eqref{eq:decay} is applied to the distance $d = |T - T^{*}|$ from the target tempo $T^{*}$ (150 or 30~BPM), but with decay rate $\lambda = 0.025$:
\begin{equation}
v = 100 \exp(-0.025\, |T - T^{*}|).
\end{equation}
With the same $\lambda = 0.1$ as the other distance-type functions, even a perceptually near-identical difference of about 5~BPM would drop the value to about 61, leaving the distance granularity out of balance with the other features. We therefore adopted the four-times-gentler $\lambda = 0.025$, consistent with the discrimination threshold of tempo perception (roughly 5\%); a 5~BPM difference then gives $v \approx 88$ and a 20~BPM difference $v \approx 61$.

\subsection*{B.5 Duration (\vflabel{60~s duration}, \vflabel{5~s duration})}

The playing time $L$ of a piece (in seconds) is obtained as the total playback time of all events in the MIDI file expanded to real time, taking tempo changes into account. Eq.~\eqref{eq:decay} is applied to the distance $d = |L - L^{*}|$ from the target duration $L^{*}$ (60 or 5 seconds), with $\lambda = 0.1$ (i.e., each second of difference is penalized as much as one semitone):
\begin{equation}
v = 100 \exp(-0.1\, |L - L^{*}|).
\end{equation}

\subsection*{B.6 Pitch class (\vflabel{A locrian PCs}, \vflabel{B major PCs})}

The target state is that the pitch classes of all notes are contained in the tone set of a specific scale. The target set $S$ is the set of pitch classes of the seven scale tones: $S = \{\text{A, A\#, C, D, D\#, F, G}\}$ for A Locrian and $S = \{\text{B, C\#, D\#, E, F\#, G\#, A\#}\}$ for B major (enharmonic equivalents are normalized to sharp spelling before matching). Writing the pitch class of each note $p_i$ as $\mathrm{pc}(p_i)$, the value is the proportion of in-set notes:
\begin{equation}
v = 100 \cdot \frac{|\{i \mid \mathrm{pc}(p_i) \in S\}|}{N}.
\end{equation}

\begin{center}
\begin{minipage}[t]{0.42\linewidth}\centering
\includegraphics[width=\linewidth]{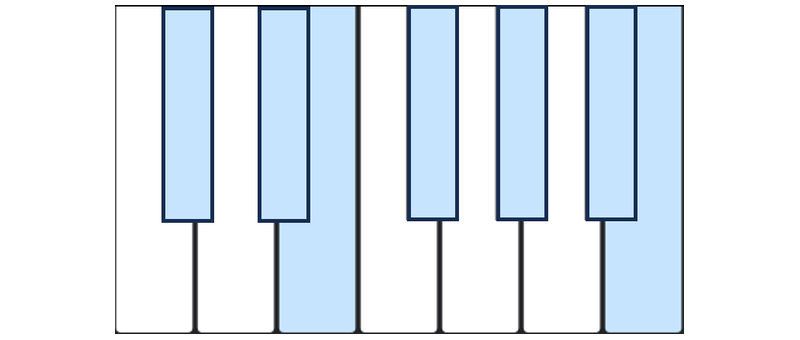}\\
{\small \vflabel{A locrian PCs}}
\end{minipage}\hspace{0.06\linewidth}
\begin{minipage}[t]{0.42\linewidth}\centering
\includegraphics[width=\linewidth]{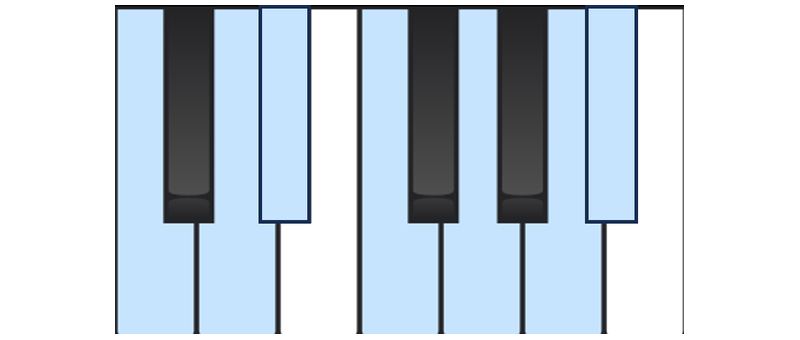}\\
{\small \vflabel{B major PCs}}
\end{minipage}
\end{center}

\subsection*{B.7 Emotion (\vflabel{M2E-Sad})}

As described in Section~4 of the main text, \vflabel{M2E-Sad} alone computes its value from the acoustic signal. The MIDI file is rendered to WAV with FluidSynth (FluidR3\_GM SoundFont, 44.1~kHz sampling rate) and fed to the pretrained Music2Emo model. Music2Emo is a multi-task model based on MERT-v1-95M that outputs a probability for each of the 56 mood/theme tags of the MTG-Jamendo dataset. The value uses the sigmoid probability $P_{\mathrm{sad}} \in [0, 1]$ of the \texttt{sad} tag:
\begin{equation}
v = 100 \cdot P_{\mathrm{sad}}.
\end{equation}

\suppappendix{Full Tables for Experiment 2}

\setcounter{table}{0}
\renewcommand{\thetable}{C\arabic{table}}

This appendix provides the full numerical tables for the Experiment~2 analyses summarized in Section~6 of the main text. Table layouts and symbols follow the definitions in the main text.

\subsection*{C.1 Prediction Task and Diagnostics}

\begin{table}[!ht]
\centering
\footnotesize
\resizebox{\linewidth}{!}{%
\begin{tabular}{lccccc}
\hline
Value function & (i) & (ii) & (iii) & (iv) & (v) \\
\hline
\vflabel{C2--B2 range}
& 0.489 & $0.509\pm0.040$ & $0.488\pm0.047$ & $0.524\pm0.037$ & $0.498\pm0.045$ \\
\vflabel{Descending contour}
& 0.553 & $0.535\pm0.062$ & $0.563\pm0.089$ & $0.532\pm0.073$ & $0.589\pm0.066$ \\
\vflabel{60~s duration}
& 0.532 & $0.510\pm0.037$ & $0.523\pm0.033$ & $0.525\pm0.029$ & $0.546\pm0.023$ \\
\vflabel{M2E-Sad}
& 0.525 & $0.488\pm0.055$ & $0.459\pm0.039$ & $0.480\pm0.057$ & $0.456\pm0.025$ \\
\vfcomb{Ascending contour}{150~BPM tempo}
& 0.508 & $0.543\pm0.037$ & $0.586\pm0.034$ & $0.556\pm0.031$ & $0.615\pm0.032$ \\
\hline
\end{tabular}
}
\caption{Per-value-function accuracy in the prediction task (numerical table corresponding to the figure in the main text). Condition (i) shows the values on the common prediction set; conditions (ii)--(v) show mean $\pm$ standard deviation over the 30 initial pieces.}
\label{tab:supp_exp2_prediction_accuracy}
\end{table}

\begin{table}[!ht]
\centering
\small
\begin{tabular}{lrr}
\hline
Value function & Pearson's $r$ & Spearman's $\rho$ \\
\hline
\vflabel{C2--B2 range} & $-0.267$ & $-0.240$ \\
\vflabel{Descending contour} & $0.126$ & $0.115$ \\
\vflabel{60~s duration} & $0.207$ & $0.216$ \\
\vflabel{M2E-Sad} & $0.260$ & $0.313$ \\
\vfcomb{Ascending contour}{150~BPM tempo} & $-0.276$ & $-0.185$ \\
\hline
\end{tabular}
\caption{Correlations between the final adaptation in Experiment~1 and the accuracy of condition (iv) ($n=30$ per value function). None is significant after Holm correction.}
\label{tab:supp_exp2_adaptation_prediction_correlation}
\end{table}

The final adaptation ratio and condition (iv)'s accuracy were correlated across the 30 initial pieces separately for each value function. None of the Pearson or Spearman correlations was significant after Holm correction, and their directions were inconsistent across functions. Thus, the magnitude of within-loop adaptation did not predict preference prediction on unseen music.

\begin{table}[!ht]
\centering
\small
\begin{tabular}{lrrrr}
\hline
Value function
& $\rho$
& Bottom 25\%
& Top 25\%
& Difference \\
\hline
\vflabel{C2--B2 range} & 0.220 & 0.453 & 0.584 & $+0.131$ \\
\vflabel{Descending contour} & 0.106 & 0.510 & 0.549 & $+0.039$ \\
\vflabel{60~s duration} & $-0.119$ & 0.535 & 0.441 & $-0.095$ \\
\vflabel{M2E-Sad} & $-0.098$ & 0.513 & 0.453 & $-0.059$ \\
\vfcomb{Ascending contour}{150~BPM tempo} & 0.192 & 0.468 & 0.598 & $+0.131$ \\
\hline
\end{tabular}
\caption{Relation between the value difference $\Delta v$ and accuracy in condition (iv). $\rho$ is the rank correlation between $\Delta v$ and accuracy; bottom and top 25\% are the accuracies of the pair groups split by quartiles of $\Delta v$.}
\label{tab:supp_exp2_value_gap_accuracy}
\end{table}

If the LLM used criteria aligned with the value function, pairs separated by a larger $\Delta v$ should be easier to order. The weak negative relations for \vflabel{60~s duration} and \vflabel{M2E-Sad} are difficult to explain by random judgments alone, which should remain near chance as $\Delta v$ changes. They suggest that, for more widely separated pairs, the proxy cues used by the LLM became less consistent with the value-function ordering. By contrast, the two clearest positive relations occurred for \vflabel{C2--B2 range} and \vfcomb{Ascending contour}{150~BPM tempo}, the two functions for which condition (iv) significantly exceeded condition (i). These observations are descriptive and do not identify the judgment mechanism.

\begin{table}[!ht]
\centering
\small
\begin{tabular}{lr}
\hline
Condition & Disagreement rate \\
\hline
(i) & 40.92\% \\
(ii) & 31.40\% \\
(iii) & 35.37\% \\
(iv) & 25.24\% \\
(v) & 37.85\% \\
\hline
\end{tabular}
\caption{Disagreement rate of choices when the presentation order was reversed.}
\label{tab:supp_exp2_order_disagreement}
\end{table}

As a supplementary observation, restricting attention to the pairs whose choice in condition (i) changed with presentation order, an average of 66.6\% of them changed to choosing the same piece in condition (ii), 63.0\% in condition (iii), 72.8\% in condition (iv), and 61.0\% in condition (v). While condition (iv) showed the highest rate of this convergence, the mean accuracy of condition (iv) on the pairs that became consistent was 52.9\%, so the convergence did not necessarily occur in the direction correct with respect to the value function.

\subsection*{C.2 Generation Task}

\begin{table}[!ht]
\centering
\footnotesize
\resizebox{\linewidth}{!}{%
\begin{tabular}{lccccc}
\hline
Value function & (i) & (ii) & (iii) & (iv) & (v) \\
\hline
\vflabel{C2--B2 range}
& 83.70 & $85.33\pm2.67$ & $85.84\pm2.42$ & $87.63\pm2.34$ & $83.02\pm2.13$ \\
\vflabel{Descending contour}
& 49.55 & $41.96\pm13.46$ & $44.78\pm6.93$ & $48.76\pm10.80$ & $48.72\pm7.32$ \\
\vflabel{60~s duration}
& 6.62 & $13.03\pm7.90$ & $23.99\pm14.28$ & $24.39\pm15.73$ & $18.63\pm10.47$ \\
\vflabel{M2E-Sad}
& 66.14 & $61.37\pm13.86$ & $65.05\pm9.51$ & $70.87\pm11.86$ & $74.02\pm11.81$ \\
\vfcomb{Ascending contour}{150~BPM tempo}
& 48.39 & $51.99\pm5.91$ & $52.70\pm6.16$ & $56.09\pm6.27$ & $42.45\pm9.25$ \\
\hline
\end{tabular}
}
\caption{Per-value-function mean value of the generated music in the generation task (numerical table corresponding to the figure in the main text). Condition (i) shows the values of the common 30 pieces; conditions (ii)--(v) show mean $\pm$ standard deviation over the 30 initial pieces.}
\label{tab:supp_exp2_generation_mean}
\end{table}

For the within-cell maximum rather than the mean, condition (i) was highest for \vflabel{60~s duration} and \vflabel{Descending contour} (99.73 and 75.00, respectively). Thus, the acquired context raised average value, while context-free generation could still occasionally reach a high value in its single best piece.

\subsection*{C.3 Acoustic Proximity and Relation to Experiment-1 Adaptation}

\begin{table}[!ht]
\centering
\small
\begin{tabular}{lrrrr}
\hline
Value function & (ii) & (iii) & (iv) & (v) \\
\hline
\vflabel{C2--B2 range} & 0.055 & 0.034 & 0.025 & 0.035 \\
\vflabel{Descending contour} & 0.054 & 0.033 & 0.030 & 0.032 \\
\vflabel{60~s duration} & 0.067 & 0.030 & 0.032 & 0.035 \\
\vflabel{M2E-Sad} & 0.055 & 0.032 & 0.031 & 0.033 \\
\vflabel{Ascending contour}~$\times$~\vflabel{150~BPM tempo} & 0.055 & 0.033 & 0.034 & 0.033 \\
\hline
\end{tabular}
\caption{Minimum MERT distance $d_{\min}$ from generated pieces to the associated analysis reference set (mean over the 30 initial pieces). The reference set is the Experiment-1 generations of the cell for conditions (ii)--(iv) and the 60 diverse-generation pieces for condition (v); comparisons are valid only between conditions sharing the same reference set. In condition (ii), the reference pieces were not included in the LLM prompt.}
\label{tab:supp_exp2_mert_distance}
\end{table}

For the history-containing conditions (iii) and (iv), $d_{\min}$ was below even the nearest-neighbor distance within the set of 30 initial pieces (mean 0.051). The mean distance to the reference set, meanwhile, matched typical between-piece levels, and the maximum distance was nearly constant across conditions. In condition (iv), generations closer to the reference set had higher value within a cell (cell-mean Spearman $\rho=-0.13$ to $-0.20$, Holm-significant for all five functions). The mean-value gain over condition (ii) was also larger in cells where adding history shrank $d_{\min}$ more (Pearson $r=0.47$--$0.51$, Holm-significant for 3 of 5 functions). By contrast, \vfcomb{Ascending contour}{150~BPM tempo} showed a significant value gain uncorrelated with the approach toward referenced pieces. These results are correlational and do not establish causality between proximity and value improvement.

We also correlated, per initial piece, the final adaptation ratio in Experiment~1 with the mean value of generated music. After Holm correction, only condition (iii) for \vflabel{C2--B2 range} was significant (Pearson $r=0.617$). Uncorrected positive tendencies concentrated in the history-containing conditions and were nearly absent in condition (v); evidence that cells adapting well in Experiment~1 transferred more strongly in generation was therefore limited.

\clearpage
\suppappendix{Illustrative Case Analyses from Experiment 1}

\setcounter{table}{0}
\renewcommand{\thetable}{D\arabic{table}}
\setcounter{figure}{0}
\renewcommand{\thefigure}{D\arabic{figure}}

\begin{figure}[b]
\centering
\includegraphics[width=\textwidth]{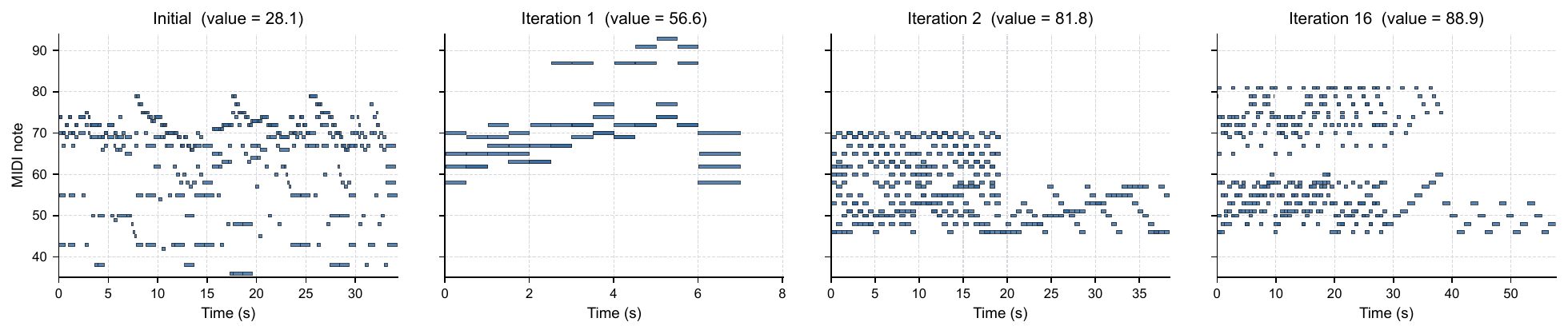}
\caption{Reference-piece trajectory for \vflabel{M2E-Sad} with \texttt{music20}: piano rolls of the initial piece and iterations 1, 2, and 16.}
\label{fig:supp_exp1_sad}
\end{figure}

This appendix provides the full analyses of four cases used to illustrate how hypotheses and musical changes led to higher values in Experiment~1. The cases represent qualitatively different adaptation processes and are not statistically representative. Table~\ref{tab:supp_exp1_cases} reports the iteration-level changes and inferred value criteria. The two \vflabel{Ascending contour} trajectories are visualized in Figure~6 of the main article, and Figure~\ref{fig:supp_exp1_sad} shows the \vflabel{M2E-Sad} trajectory.

The cases illustrate stepwise exploration (\vflabel{150~BPM tempo}), retention and interpretation of an accidental success (\vflabel{Ascending contour}, \texttt{music7}), implicit inheritance without verbalization (\vflabel{Ascending contour}, \texttt{music10}), and acquisition of cues specific to the evaluation model (\vflabel{M2E-Sad}). In both \vflabel{Ascending contour} cases, ascending sequences emerged while the pitch content was simplified. This tendency may partly explain the asymmetry between ascending and descending contours observed in Experiment~1, although the cases do not establish a general mechanism.

\begin{table*}[p]
\centering
\small

{
\setlength{\tabcolsep}{4pt}
\renewcommand{\arraystretch}{1.22}

\newlength{\iterlabelwidth}
\settowidth{\iterlabelwidth}{iter 12: 100.0\quad}
\newcommand{\iterrow}[2]{\par\hangindent=\iterlabelwidth \hangafter=1 \noindent\makebox[\iterlabelwidth][l]{#1}#2}

\begin{tabular}{
@{}
>{\raggedright\arraybackslash}p{0.10\textwidth}
>{\raggedright\arraybackslash}p{0.44\textwidth}
>{\raggedright\arraybackslash}p{0.38\textwidth}
@{}
}
\hline
Setting
& Iterations, values, and main musical changes
& Inferred value criteria and role of the case \\
\hline

\vflabel{150~BPM tempo}\newline
\texttt{music30}\newline
(initial value: 13.9)
&
\iterrow{iter 1: 47.2}{Introduced meter changes and added a 6/8 section.}
\iterrow{iter 10: 55.8}{Accelerated the 6/8 section to 140~BPM and changed the timbre to flute.}
\iterrow{iter 17: 67.0}{Changed one measure to 150~BPM and added dynamics at the opening.}
&
Captured the importance of raising the tempo, but interpreted local tempo manipulation as what matters, rather than moving the whole piece toward 150~BPM. Grasped the direction of the target stepwise.
\\[6pt]
\hline

\vflabel{Ascending contour}\newline
\texttt{music7}\newline
(initial value: 50.3)
&
\iterrow{iter 1: 92.6}{Generated from a hypothesis aiming at a repeat structure and dotted rhythms; incidentally, the pitch content changed into a long ascending sequence.}
\iterrow{iter 20: 97.5}{Kept the sequence while adjusting dynamic transitions and accent placement.}
&
Reached a high value by chance at iteration 1, and through subsequent trial and error inferred that preserving the sequence \texttt{C D E F G A B c...} was important. Using the accidental success as a cue, inferred value criteria partially consistent with the true value function.
\\[6pt]
\hline

\vflabel{Ascending contour}\newline
\texttt{music10}\newline
(initial value: 48.4)
&
\iterrow{iter 1: 67.4}{Simplified from five voices to two, introducing an A-major scalar melody and I--V--vi--IV harmony.}
\iterrow{iter 6: 89.4}{Incidentally to a composition intended to change the rhythm, replaced the melody with a fully ascending scale.}
\iterrow{iter 9: 93.3}{Simplified the melody to one whole note per measure.}
\iterrow{iter 12: 100.0}{Changed to one melody note and one chord per measure, each followed by a rest.}
&
The value-criteria descriptions converged on A-major key, two voices, one note per measure, root-position triads, and uniform rhythm, and never verbalized the ascending direction at any recorded point. The accidentally produced ascending sequence was inherited implicitly through the carry-over of the reference piece, reaching an adaptation ratio of 100.
\\[6pt]
\hline

\vflabel{M2E-Sad}\newline
\texttt{music20}\newline
(initial value: 28.1)
&
\iterrow{iter 1: 56.6}{Simplified the melody and accompaniment.}
\iterrow{iter 2: 81.8}{Introduced functional harmony in B$\flat$ major.}
\iterrow{iter 8: 82.8}{Added call-and-response motifs \texttt{d c B A} and \texttt{e f g a}.}
\iterrow{iter 16: 88.9}{Added a short bridge oriented toward F major.}
&
The value-criteria descriptions converged on functional harmony in B$\flat$ major, a stable 4/4 meter, fixed call-and-response motifs, and a short bridge. Adapted to the model-specific musical features that Music2Emo classifies as \texttt{sad}, rather than to expressions generally associated with sadness such as minor keys or slow tempi.
\\[6pt]
\hline
\end{tabular}
}
\caption{Four cases illustrating the process of iterative adaptation and value-criteria learning}
\label{tab:supp_exp1_cases}
\end{table*}

\clearpage
\suppappendix{Implementation and Reproducibility Settings}

\setcounter{table}{0}
\renewcommand{\thetable}{E\arabic{table}}

\begin{sloppypar}

All LLM calls used \texttt{openai/gpt-oss-120b}, revision
\texttt{b5c939de\allowbreak 8f754692\allowbreak c1647ca7\allowbreak 9fbf85e8\allowbreak c1e70f8a}, with its MXFP4
weights. The model was served on NVIDIA A100 GPUs with vLLM
\texttt{0.10.2.dev2+\allowbreak gf5635d62e.d20250807}; the vLLM Docker image digest was
\texttt{sha256:\allowbreak 23c3feef\allowbreak ba723be9\allowbreak 7ff9e9bd\allowbreak 769aed7d\allowbreak 165839a7\allowbreak 9bc042eb\allowbreak 8f3a13dd\allowbreak 2a469e1c}.
The server context limit was 131,072 tokens, and tensor parallelism was 2 or 4,
depending on the A100 machine. The software and decoding configuration was not
changed on the secondary execution machine.

\begin{table}[!ht]
\centering
\scriptsize
\begin{tabular}{lrrrr}
\hline
Task & Temperature & Top-$p$ & Reasoning effort & Max. completion tokens \\
\hline
Experiment 1, proposed system & 1.0 & 1.0 & low & 16,384 \\
Experiment 1, diverse generation & 1.0 & 1.0 & low & 16,384 \\
Experiment 1, ablation & 1.0 & 1.0 & low & 16,384 \\
Experiment 2, prediction & 0.0 & 1.0 & low & 4,096 \\
Experiment 2, generation & 1.0 & 1.0 & low & 8,192 \\
\hline
\end{tabular}
\caption{LLM decoding settings. No stop sequence was supplied.}
\label{tab:supp_llm_settings}
\end{table}

We used low reasoning effort because, in preliminary trials, higher settings
more often violated ABC formatting or failed to return all three requested
hypotheses.

The base seed was 74. In Experiment~1, retries after a formatting or ABC
validation failure incremented the seed by one, with at most seven retries. In
the Experiment-2 generation task, each piece was generated by an independent
LLM call with its context fixed; invalid-output retries added 1,000 to the seed,
with at most three retries. The client allowed two retries for transport or API
errors, retaining the request seed, and used a 1,800-second timeout.

All 480 Experiment-1 cells completed iteration 20 with no exclusions. LLM
input length peaked at 116.2k tokens, within the context limit, and 17 of the
28,800 ABC generations (0.06\%) reached the conversion-validation retry limit.

Experiment~2 derived cell-specific seeds from the base seed and a textual
namespace. The original implementation used Python's process-randomized
\texttt{hash()} while \texttt{PYTHONHASHSEED} was unset; consequently, the
namespace-derived assignments cannot be reconstructed from the base seed alone
across fresh Python processes. The exact seed supplied to every reported LLM
call was therefore retained in the CSV and JSONL generation logs and will be
released with the generated data. These recorded call-level seeds, rather than
the derivation alone, define the reported runs. Even with identical prompts and
seeds, concurrent vLLM batch composition may change an individual output; the
reported analyses average over multiple generations or initial pieces to reduce
this variability.

For the acoustic-distance analyses, MERT-v1-95M revision
\texttt{12af15fe\allowbreak f9d0ac83\allowbreak 8c3f475b\allowbreak fbbf26d2\allowbreak 060dd4f5} was used. Audio was converted
to mono, resampled to 24~kHz, and represented by temporal mean pooling of the
final hidden layer. The Music2Emo evaluator used repository commit
\texttt{c5f7984f\allowbreak 78d9928b\allowbreak 841f25bd\allowbreak 05a47377\allowbreak c12c58ca} and checkpoint revision
\texttt{b036e594\allowbreak 71583c3d\allowbreak 5b30c69e\allowbreak 63e8c732\allowbreak 3cc36c4a}; the SHA-256 of
\texttt{J\_all.ckpt} was
\texttt{deaceb29\allowbreak 1f7974de\allowbreak b688167d\allowbreak 3639b7f6\allowbreak eb66eb07\allowbreak 15824668\allowbreak 61839377\allowbreak 7d1e07a5}.
Music conversion and synthesis used abc2midi 4.68 and FluidSynth 2.2.5.
Statistical analyses used SciPy 1.15.3 and statsmodels 0.14.6. The released
repository will provide the exact code snapshot, prompts reproduced in
Appendix~A, dependency lock files, call-level seeds, and analysis scripts.

\end{sloppypar}

\finishsupplement

\end{document}